%% file: 0-main.tex
\documentclass[preprint]{article}

\usepackage{microtype}
\usepackage{graphicx}
\usepackage{subfigure}
\usepackage{booktabs} 

\usepackage{icml2026}
\usepackage{xurl}

\usepackage{amsmath}
\usepackage{amssymb}
\usepackage{mathtools}
\usepackage{amsthm}

\usepackage[capitalize,noabbrev]{cleveref}

\theoremstyle{plain}

\theoremstyle{definition}

\theoremstyle{remark}

\usepackage[textsize=tiny]{todonotes}

\usepackage{xspace}
\newcommand{\oursched}{\textit{SageSched}\xspace}

\usepackage{pifont}
\usepackage{xcolor}

\newcommand{\phm}[1]{\vspace{.4em} \noindent\textbf{#1}\hspace{.5em}}

\icmltitlerunning{{SageSched}: Efficient LLM Scheduling Confronting Demand Uncertainty and Hybridity}

\begin{document}

\twocolumn[
\icmltitle{\oursched: Efficient LLM Scheduling Confronting \\ Demand Uncertainty and Hybridity}

\icmlsetsymbol{equal}{*}
\begin{icmlauthorlist}
\icmlauthor{Zhenghao Gan}{yyy}
\icmlauthor{Yichen Bao}{yyy}
\icmlauthor{Yifei Liu}{yyy}
\icmlauthor{Chen Chen}{yyy}
\icmlauthor{Quan Chen}{yyy}
\icmlauthor{Minyi Guo}{yyy}
\end{icmlauthorlist}
\icmlaffiliation{yyy}{Shanghai Jiao Tong University}
\icmlcorrespondingauthor{Chen Chen}{chen-chen@sjtu.edu.cn}
\icmlkeywords{Machine Learning, ICML}
\vskip 0.3in
]


\printAffiliationsAndNotice{}  

\begin{abstract}
    Efficient LLM inference scheduling is crucial for user experience.
    However, LLM inferences exhibit remarkable demand \emph{uncertainty} (with unknown output length beforehand) and \emph{hybridity} (being both compute and memory intensive). Existing LLM schedulers rely on simple heuristics or focus purely on compute resource, suffering suboptimal performance. 
		
    In this work, we propose \textit{SageSched}, an efficient LLM scheduler that properly handles demand uncertainty and hybridity of inference workloads.
    \textit{SageSched} combines prompt contents with the past inference results to predict output-length distribution in a light-weight and also accurate manner.
    Meanwhile, it models the true service cost of an inference request with both compute and memory aspects considered.
    Finally, \textit{SageSched} employs an uncertainty-aware scheduling policy that can yield the best overall efficiency given the request cost distributions. 
    Testbed experiments over diverse setups confirm that \textit{SageSched} can attain an efficiency improvement of over 28.7\%. 

\end{abstract}

\input{1-intro}
\input{2-background}
\input{3-solution}

\input{4-evaluation}

\input{5-related}
\input{6-conclusion}

\begingroup
\raggedright
\bibliography{main}
\bibliographystyle{icml2026}
\endgroup

\input{7-appendix}

\end{document}

%% file: 1-intro.tex
\section{Introduction}
\label{sec:intro}

Large language models (LLMs) have revolutionized diverse domains, from conversational assistants~\cite{openai_chatgpt,deepseek} to agentic and embodied systems~\cite{chen2023typefly,wang2024llm}.
With the widespread adoption of LLMs, massive LLM inference requests are often concurrently served on GPU servers~\cite{sun2024llumnix}. 
When scheduling LLM requests, it is increasingly crucial to minimize their overall service latency, or \emph{Time-to-Last-Token} (TTLT), to optimize the end-to-end user experience~\cite{shahout2024don,qiu2024efficient}.

Nonetheless, contrary to conventional workloads, LLM inferences have two challenging characteristics in resource demands: \emph{uncertainty} and \emph{hybridity}. 

On the one hand, due to the auto-regressive nature, the output token length of an LLM inference is \emph{non-deterministic} beforehand, exhibiting strong uncertainty~\cite{yu2022orca}; in contrast, OS~\cite{lozi2016linux} or big data~\cite{ousterhout2015making} workloads are often recurring with stable resource demands. 
On the other hand, with vast matrix-based operations and heavy KVCache reliance~\cite{qin2024mooncake}, LLM inferences are both \emph{compute} and \emph{memory} intensive; in contrast, with virtual memory techniques~\cite{abrossimov1989generic}, OS schedulers only care about compute cost in process scheduling.

Confronting the demand uncertainty and hybridity of LLM inference workloads, existing LLM schedulers fail to attain high efficiency. 
Mainstream LLM frameworks like vLLM~\cite{kwon2023efficient} and SGLang~\cite{zheng2024sglang} adopt the classical \emph{first-come-first-serve} policy, which often inflates the overall TTLT due to \emph{head-of-line-blocking}. 
While some recent schedulers~\cite{shahout2024don,qiu2024efficient,fu2024efficient} seek to approximate the \emph{shortest-job-first} policy with the predicted output length, they are still suboptimal. 
{First}, those schedulers commonly rely on a fine-tuned model to directly predict the inference output-length, which is both \emph{heavy-weight} (due to costly training and predicting) and \emph{inaccurate} (due to the difficulty to emulate the original generation effect).
Meanwhile, existing schedulers queue the pending requests based on their output token length, which ignores demand hybridity and fails to capture their true service costs.
Moreover, those schedulers commonly use a single (expected mean) value as the queuing index, failing to leverage the inherent uncertainty distribution information for the best queuing effect.

In this paper, we design \oursched, an LLM scheduler that attains high efficiency by properly handling the demand uncertainty and hybridity characteristics of inference workloads.
There are three key techniques in \oursched.
\emph{First}, in demand prediction, motivated by the correlation between prompt similarity and output-length similarity, we propose \emph{semantic-aware history-based predictor}, which combines the \emph{prompt contents} with the \emph{past inference results} to predict an output-length distribution.
In this way, we can avoid the burden to maintain a heavy-weight prediction model for each LLM or the complexity to directly emulate the original generation effect. 
\emph{Second}, in cost modeling, we consider resource contention in both \emph{compute} and \emph{memory} aspects. 
By respectively studying LLM serving process in compute- and memory-bound scenarios, we build a \emph{unified cost model} that can always capture the true service cost of an LLM inference.
\emph{Third}, given that the service cost of an inference is best depicted as a distribution, we adopt an \emph{uncertainty-aware request scheduling policy}.
Specifically, when queuing each request, we calculate the \emph{Gittins index} out of its cost distribution, which is known to yield the best overall performance for \emph{jobs with unknown durations but known duration-distributions}~\cite{gittins1979bandit}; we also periodically refresh ongoing inferences' Gittins indices to ensure timeliness of the scheduling plan.

We have implemented \oursched on top of the popular vLLM framework~\cite{vllm_v0_8_2}, and further evaluated its performance with diverse real-world LLM traces. 
Our experiments show that \oursched surpasses state-of-the-art LLM schedulers by over 28.7\% in terms of the average TTLT. 
Besides, our deep dive analysis further confirm the superiority of each (prediction, modeling and scheduling) techniques adopted in \oursched. 
Moreover, large-scale simulations confirm that \oursched can scale well in large clusters with relatively low scheduling overheads.

In summary, this paper makes the following contributions:
\begin{itemize}
    \item We identify with testbed measurements the limitations of existing schedulers when confronting the uncertainty and hybridity characteristics of LLM workloads. 
    \item We design \oursched, a TTLT-efficient LLM scheduler integrating three key techniques: semantic-aware history-based prediction, resource-bound-based cost modeling, and uncertainty-aware requesting queuing. 
    \item We evaluate \oursched performance with both testbed experiments and simulations, which show that it can attain remarkable efficiency enhancement, with a TTLT improvement of over 28.7\%.

\end{itemize}

%% file: 2-background.tex
\section{Background and Motivation}
\label{sec:background}

\subsection{The LLM Scheduling Problem}
\label{subsec:problem}

\begin{figure}[t]
    \centering
    \subfigure[Output length variations of ten randomly sampled prompts from SharedGPT, each collected over 100 inference trial runs.]{
        \includegraphics[width=0.45\linewidth]{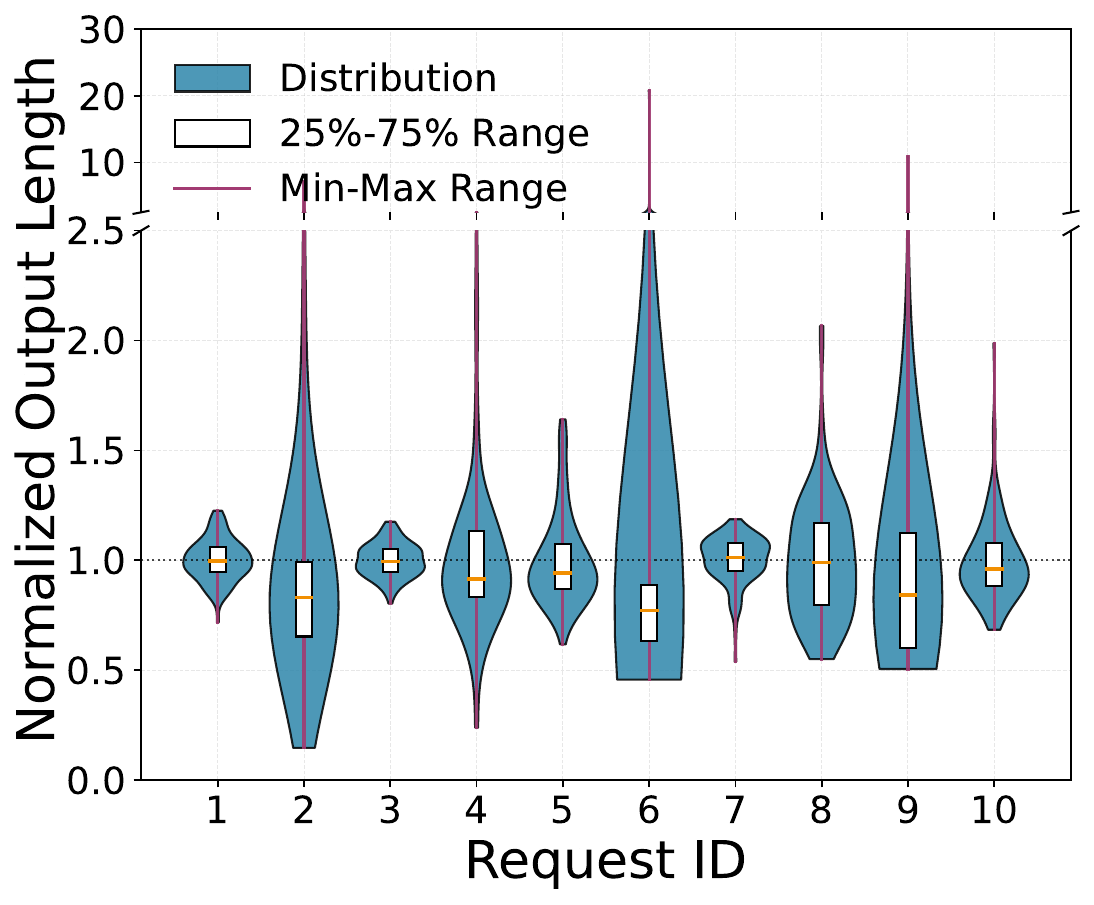}
        \label{fig:background_uncertainty}
    } 
    \hfill
    \subfigure[Scatterplot (execution time, peak memory usage) of requests from three datasets, each profiled alone with an H800 GPU.]{
        \includegraphics[width=0.45\linewidth]{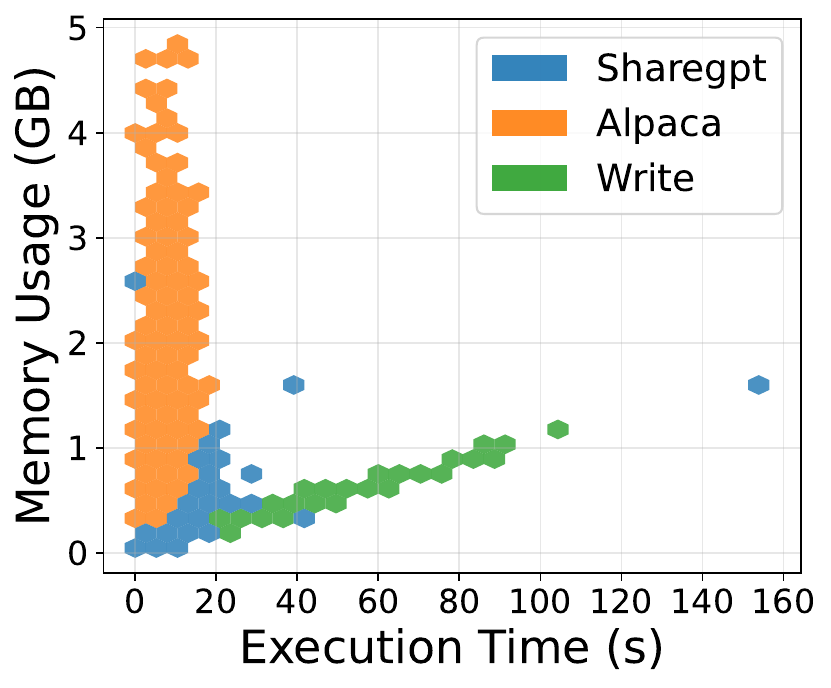}
        \label{fig:background_hybridity}
    }
    \caption{Empirical evidences on the \emph{uncertainty} and \emph{hybridity} characteristics of LLM inferences' resource demands.}

    \label{fig:background_uncertainty_and_hybridity}
\end{figure}

\phm{Demand \emph{uncertainty} and \emph{hybridity} of LLM inference workloads.}
Large language models (LLMs)~\cite{deepseek, llama_official} have nowadays been widely adopted in diverse scenarios like chatbot conversation~\cite{openai_chatgpt}, code generation~\cite{cursor_ai} and embodied intelligence~\cite{wang2023voyager}.
Notably, LLM inferences are conducted in an auto-regressive manner: given a prompt input, each inference task continuously generates new tokens based on its latest token sequence so far---until a special \texttt{<EOS>} token is yielded~\cite{yu2022orca}.
As shown in Fig.~\ref{fig:background_uncertainty}, even a fixed prompt would generate different\footnote{In this paper, we use a temperature of 0.6 in all the inferences, a default setup for many popular LLMs~\cite{llama3_1_8b_instruct,qwen3_32b}.}
output tokens in different runs.
In that sense, the ultimate output length of an inference request can only be known after inference completion, exhibiting strong \emph{uncertainty}.

Apart from demand {uncertainty}, demand \emph{hybridity} is also a distinct property of LLM inference workloads. 
Specifically, during the inference process, each inference commonly caches its token sequence in GPU memory (which is called KVCache~\cite{qin2024mooncake}) to avoid the recomputation overheads; provisioning such a memory space is a prerequisite for inferences to proceed in mainstream LLM frameworks like vLLM~\cite{vllm_v0_8_2}.
In Fig.~\ref{fig:background_hybridity}, we profile the memory usage and execution time of all the requests respectively from three typical LLM datasets (\emph{as selected} by \cite{hu2024inference})---\emph{SharedGPT}~\cite{sharegpt_vicuna_unfiltered_2025}, \emph{Alpaca-Summarization}~\cite{alpaca_pubmed_summarization_2025}, and \emph{{Document-Write}}~\cite{write_doc_sft_v1_2025}.
It shows both \emph{computation} and \emph{memory} resources are first-class citizens in LLM serving; in contrast, conventional OS workloads are scheduled dominantly based on the computation cost~\cite{cpu_scheduling_comparison_gfg_2026}. 

\phm{Scheduling is crucial on LLM service backends.} 
Given the booming adoption of LLMs, the LLM inference requests from diverse users are often served in a shared backend, where different requests compete for the limited compute and memory resources~\cite{Recasens2025Mind}. 
For good user experience, it is crucial to properly schedule the LLM inferences to achieve low overall latency. 
Specifically, while there are diverse latency metrics for LLM inferences: time-to-first-token (TTFT), time-per-output-token (TPOT), and time-to-last-token (TTLT), in this paper we primarily focus on TTLT due to its comprehensiveness\footnote{
First, TTLT is increasingly significant for many scenarios like agentic AI: for example, for  LLM-powered robotic control~\cite{chen2023typefly,wang2024llm}, downstream operations can only be launched after LLM generation completion. 
Second, optimizing TTLT often also helps to reduce the other metrics.
For TTFT, by avoiding head-of-line-blocking, TTLT-friendly schedulers can naturally improve TTFT (as confirmed by our later evaluation);
for TPOT---which in statistical analysis is often defined as \emph{dividing TTLT by the output token length}~\cite{wu2023fast,fu2024efficient}---reducing TTLT would proportionally improve TPOT.
}, as in a series of existing works~\cite{shahout2024don,qiu2024efficient}. 

\subsection{Limitations of Existing LLM Schedulers}
\label{subsec:limitations}

In the literature, a series of schedulers have been proposed for LLM inference workloads.
Confronting demand uncertainty, early-stage LLM schedulers usually employ \emph{demand-agnostic} scheduling heuristics. 
Production LLM frameworks like vLLM~\cite{kwon2023efficient} and SGLang~\cite{zheng2024sglang} employ a simple \emph{first-come-first-serve} (FCFS) scheduler, which suffers the head-of-line-blocking problem.
To address that problem, FastServe~\cite{wu2023fast} adopts a \emph{multi-level feedback queue} (MLFQ) to periodically degrade ongoing inferences to lower priorities after fixed service quantums---essentially approximating (yet not as good as) \emph{shortest-remaining-processing-time} (SRPT) at the cost of additional preemption overheads.

Noticing the deficiencies of \emph{demand-agnostic} schedulers, recent LLM schedulers adopt a \emph{prediction-based} methodology.
For example, the \emph{Speculative Shortest Job First} (SSJF)~\cite{qiu2024efficient} scheduler approximates \emph{shortest-job-first} (SJF) by using a fine-tuned BERT-base model, which predicts the output token length given the inference prompt.
In the meantime, Fu~\emph{et~al.}~\cite{fu2024efficient} also approximate SJF using an OPT-125M model, which predicts the requests' relative output-length \emph{rank} (rather than the exact lengths).
Recently, the TRAIL work~\cite{shahout2024don} approximates SPRT with an MLP model, which predicts the output length continuously at iteration granularity---using the feature embeddings of certain LLM layers.  

\begin{figure}[t]
    \centering
    \subfigure[Predicting a single (bucket) value fails to cover all output-length possibilities; each bucket represents an output-length range of {100} tokens.]{ 
        \includegraphics[width=0.53\linewidth]{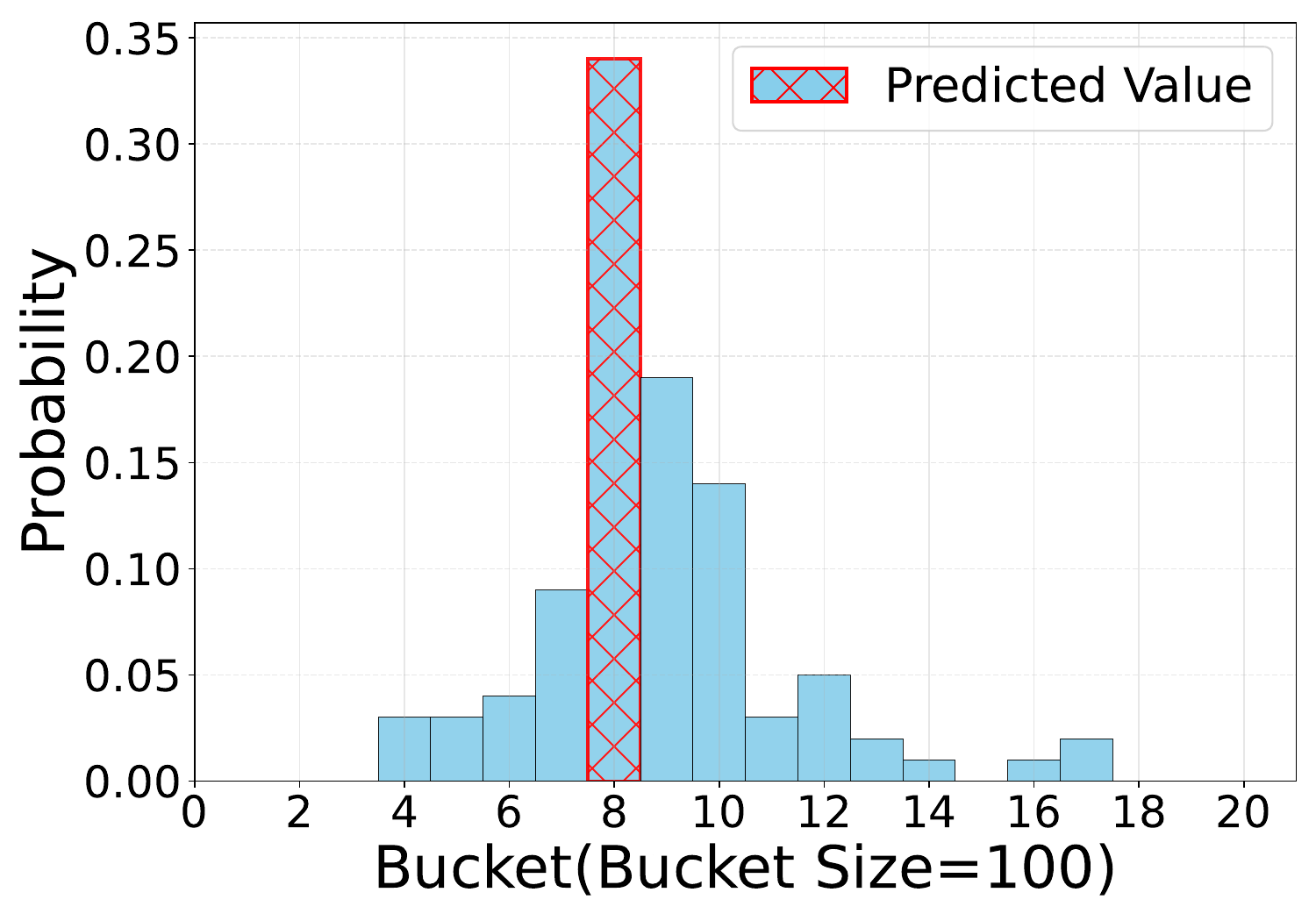}
        \label{fig:low_prediction_accuracy}
    }
    \hfill
    \subfigure[Due to KVCache constraints, prioritizing requests with shorter outputs may be suboptimal.]{ 
        \includegraphics[width=0.37\linewidth]{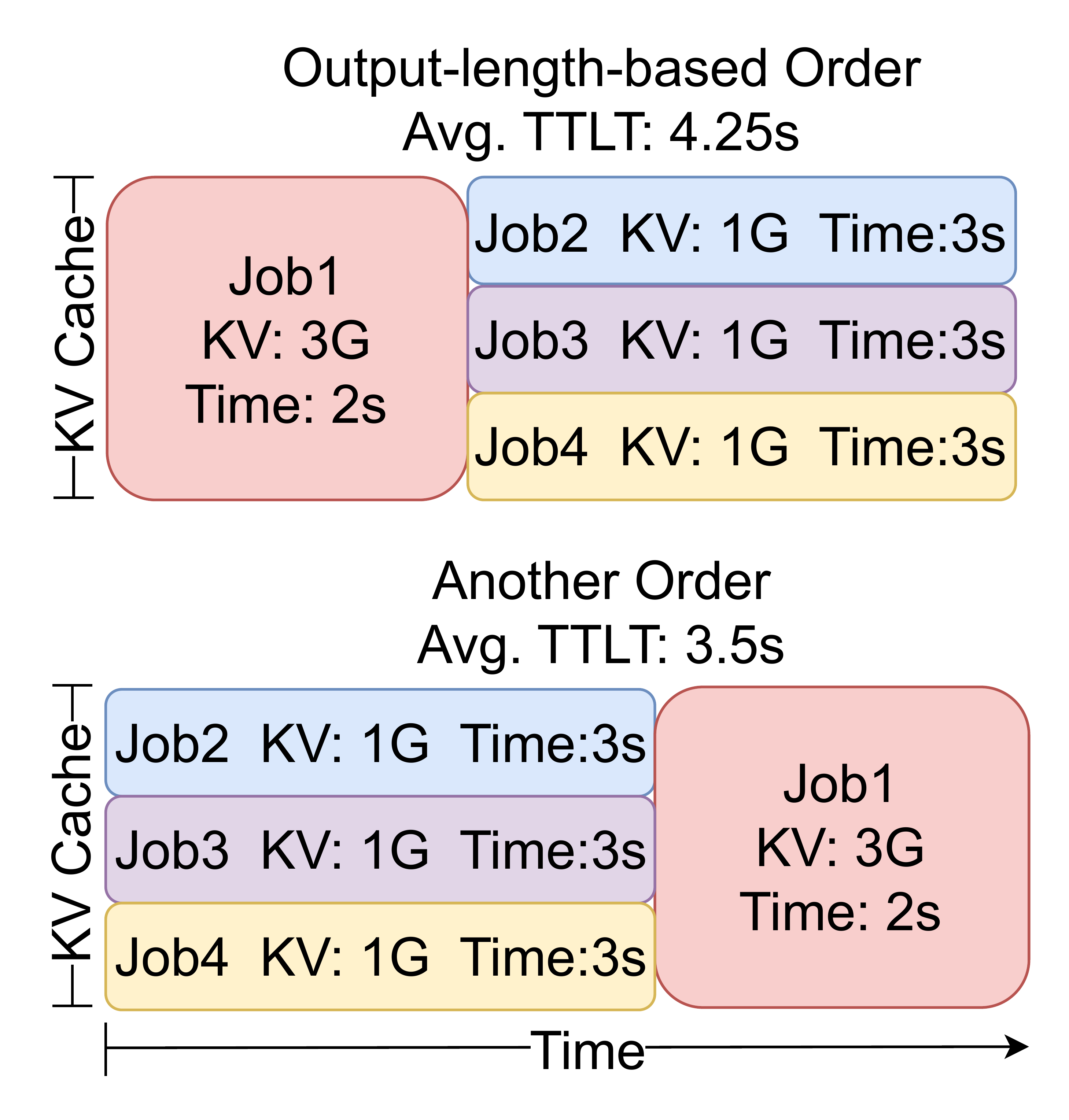}
        \label{fig:wrong_scheduling_order}
    }
    \caption{Examples elaborating the deficiencies of existing schedulers when confronting demand uncertainty and hybridity.} 
    \label{fig:deficiencies}
\end{figure}

However, the above prediction-based schedulers still fall short in handling demand \emph{uncertainty}.
First, their length predictors commonly require heavy-weight offline training (fine-tuning typically requires collecting and processing 10k+ training examples~\cite{jin2023s, fu2024efficient}), and the model trained for one LLM cannot be directly applied for another.
Second, those output-length predictors yield a \emph{single} prediction value, losing valuable probabilistic distribution information.
In fact, demand uncertainty is a built-in property of LLM inferences, and a single prediction value would hardly be accurate in reality. 
As shown in Fig.~\ref{fig:low_prediction_accuracy}, 
the single value predicted by DistillBert as in \cite{qiu2024efficient} only achieves an accuracy of 34.1\% (by hitting only one out of the totally {20} buckets).
As we show later in Sec.~\ref{subsec:gittins}, preserving the length distribution can help to make better scheduling decisions. 
Therefore, in handling demand uncertainty, we need to design a \emph{training-free} method that directly predicts the output-length \emph{distribution}.

Meanwhile, we note that those schedulers also fall short in handling demand \emph{hybridity}. 
They quantify an LLM request's service cost merely as its computation time (represented by its output token length), being inaccurate by neglecting its resource consumption of GPU memory. 
In fact, since both the input and output tokens take up KVCache spaces during inference, two inferences with the same output length may exhibit substantially different memory costs.  
As shown in Fig.~\ref{fig:wrong_scheduling_order},  when the service backend is bottlenecked on GPU memory, simply prioritizing the LLM inferences with shorter output lengths may in fact be suboptimal. 
Therefore, to handle demand hybridity of LLM inferences, we need to jointly consider their resource consumptions in both compute and memory dimensions.

To summarize, due to demand uncertainty and hybridity, existing LLM schedulers fail to achieve high overall efficiency. 
We next devise a novel scheduler for efficient LLM scheduling, with optimizations respectively in output-length prediction, cost modeling and request queuing. 

%% file: 3-solution.tex
\section{SageSched Design}
\label{sec:design}

\begin{figure}
    \centering
    \includegraphics[width=0.8\linewidth]{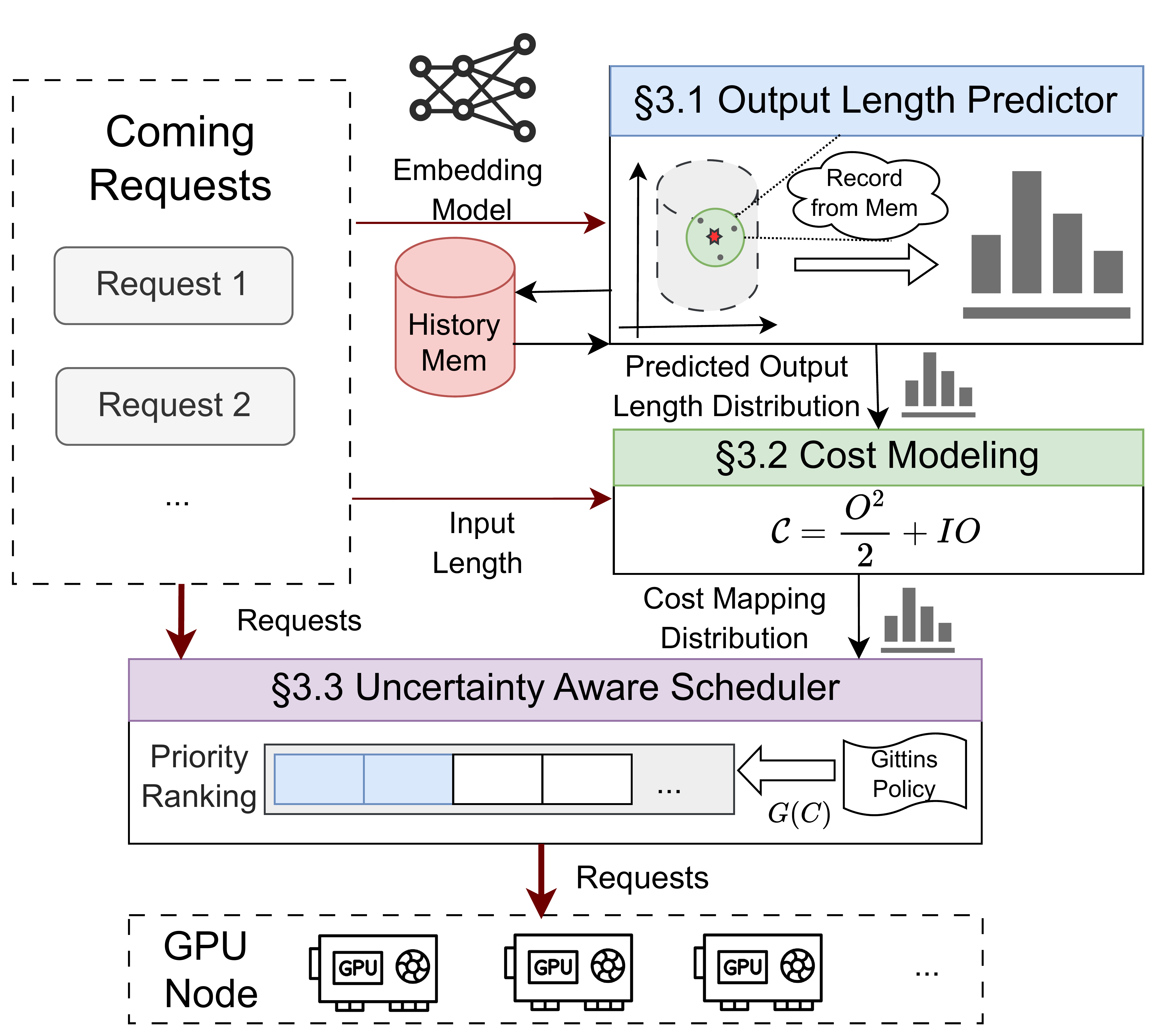}
    \caption{Overview of \oursched workflow.}
    \label{fig:overview}
\end{figure}

In this paper, we design \oursched, a novel LLM scheduler that, by properly handling dynamicity uncertainty and hybridity, attains efficient TTLT performance. 
The overall workflow of \oursched is elaborated in Fig.~\ref{fig:overview}.
For each incoming LLM inference request, \oursched first predicts its output-length distribution with a light-weight yet also accurate predictor (Sec.~\ref{subsec:predictor}).
It then quantifies the overall service cost of that request based on the predicted output length, with both the memory and compute resources considered (Sec.~\ref{subsec:cost}).
Finally, \oursched employs an uncertainty-aware scheduling algorithm to determine the runtime request queuing order, which can yield the optimal efficiency (Sec.~\ref{subsec:gittins}).

\subsection{Semantic-aware History-based Predictor}
\label{subsec:predictor}

As elaborated in Sec.~\ref{subsec:limitations}, given the built-in uncertainty of LLM decoding process, a truly accurate prediction method must yield a \emph{distribution} rather than a \emph{number} when depicting the predicted output token length.
In that regard, a straightforward solution is to customize the output-projection layers in existing prediction models (e.g., DistillBert~\cite{distilbert_docs}) for distribution prediction; such a customized model essentially seeks to---given the semantic input prompt---directly \emph{approximate} the ultimate generation effect. 
Yet, maintaining such a prediction model is still \emph{costly} (due to the training overheads) and \emph{inaccurate} (due to the complexity to approximate the inference effect).

In fact, given an inference prompt, we find its output length should be better predicted by \emph{referring to similar requests in the recent history}---instead of by directly emulating the inner generation process with a fine-tuned model.
Given the complexity of auto-regressive generation, it would be much easier to \emph{evaluate prompt similarity} than to \emph{directly emulate the generation effect}. 
Meanwhile, existing works~\cite{gong2025past} have revealed that, \emph{the overall distributions of request output length within adjacent sliding time windows are relatively stable}, indicating high reference value of the recent serving history. 
Therefore, by selecting the past requests with a similar prompt, we can potentially make \emph{easier-yet-more-accurate} demand prediction.

\begin{figure}[t]
    \centering
    \subfigure[Prompt-1 sampled from the \emph{Alpaca} dataset]{
        \includegraphics[width=0.9\linewidth]{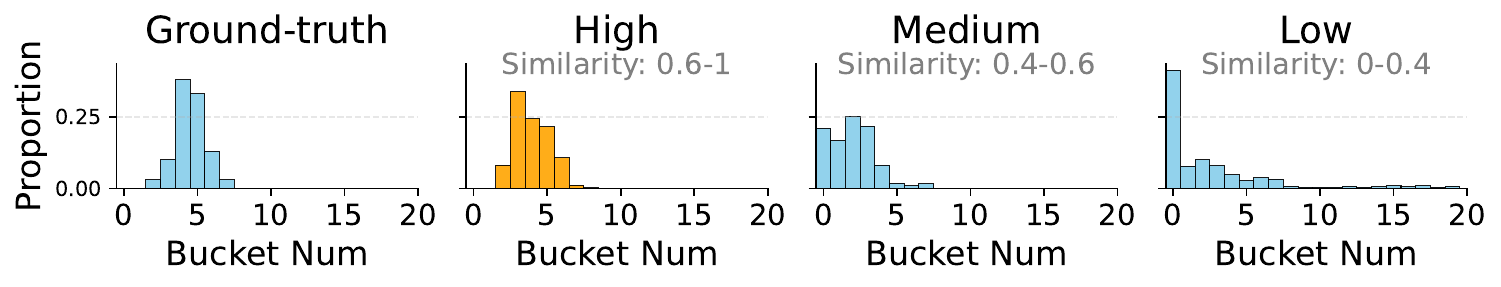}
        \label{fig:correlation-1}
    }
    \hfill
    \subfigure[Prompt-2 sampled from the \emph{Write} dataset]{
        \includegraphics[width=0.9\linewidth]{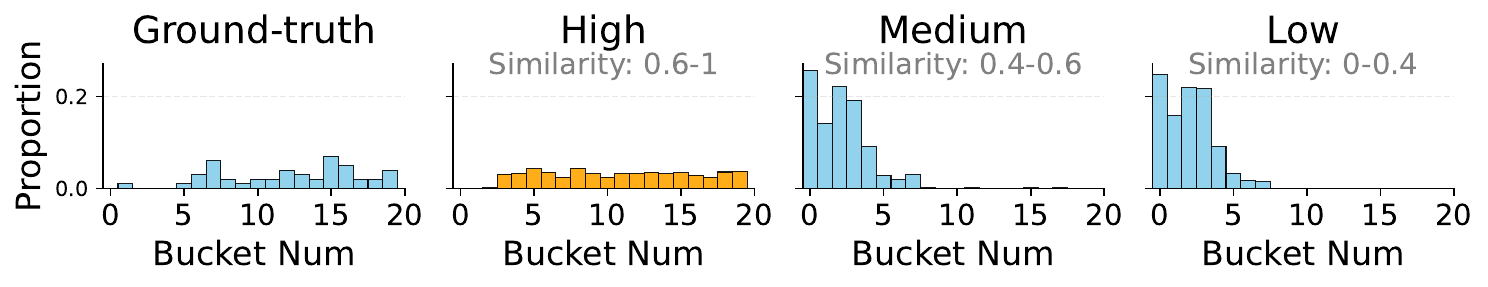}
        \label{fig:correlation-2}
    }
    \caption{Output-length distribution of a request can be better approximated by historical requests with a higher prompt similarity.} 
    \label{fig:prompt-correlation}
\end{figure}

To explore, we empirically study the correlation between prompt similarity and output-length similarity. 
We sample two prompts respectively from the Alpaca and Write datasets (elaborated in Sec.~\ref{subsec:problem}). 
For each prompt, we first serve it 100 times, obtaining an output-length distribution as the ground-truth to predict.
Then we classify the historical requests into three categories based on their similarity to the selected prompt; here similarity is quantified as the \emph{cosine similarity} between two requests' prompt embeddings. 
As shown in Fig.~\ref{fig:prompt-correlation}, for each case, there exists a clear correlation between the prompt similarity and the output-length-distribution similarity: the curve shapes of \emph{higher-similarity} buckets are \emph{closer} to the target distribution.

Given the above analysis, we propose \emph{semantic-aware history-based predictor}, which combines the \emph{prompt contents} with the \emph{historical execution status} to directly predict the output length distribution. 
Specifically, we record the inference information (prompt contents and output lengths) of recently-served requests.
For each incoming request, we search\footnote{Our history window has a size of 10,000 records and keeps updating in a \emph{FIFO} manner. 
    In cases where the high-similarity requests are insufficient (e.g., in the warming-up phase), we augment the searching set with the requests from public datasets.
    We use the efficient \emph{FAISS IndexFlat}~\cite{faiss_meta} {tool} to perform embedding search, which in general takes {less than 1 ms}.
}
for those requests with a sufficiently-similar prompt (with a similarity threshold defaulted to 0.8), and use their output-length distribution as the prediction result.
Such a method does not require fine-tuning any model, thereby being generally applicable to diverse LLMs; neither does it require emulating the complex generation process.
Therefore, in this way we can make light-weight and also accurate prediction for output-length distribution. 

\subsection{Resource-bound-based Cost Modeling}
\label{subsec:cost}

With the above prediction method, we can now obtain the output-length distribution for each upcoming request. 
Nonetheless, as explained in Sec.~\ref{subsec:limitations}, due to demand hybridity, it is deficient to directly use the output token length as the service cost of an LLM request.
In this part we need to find a comprehensive cost-depicting model with both compute and memory resources considered. 

\begin{figure}[t]
    \centering
    \subfigure[Relationship between GPU utilization and KVCache occupation as batch size increases.
    ]{
        \includegraphics[width=0.45\linewidth]{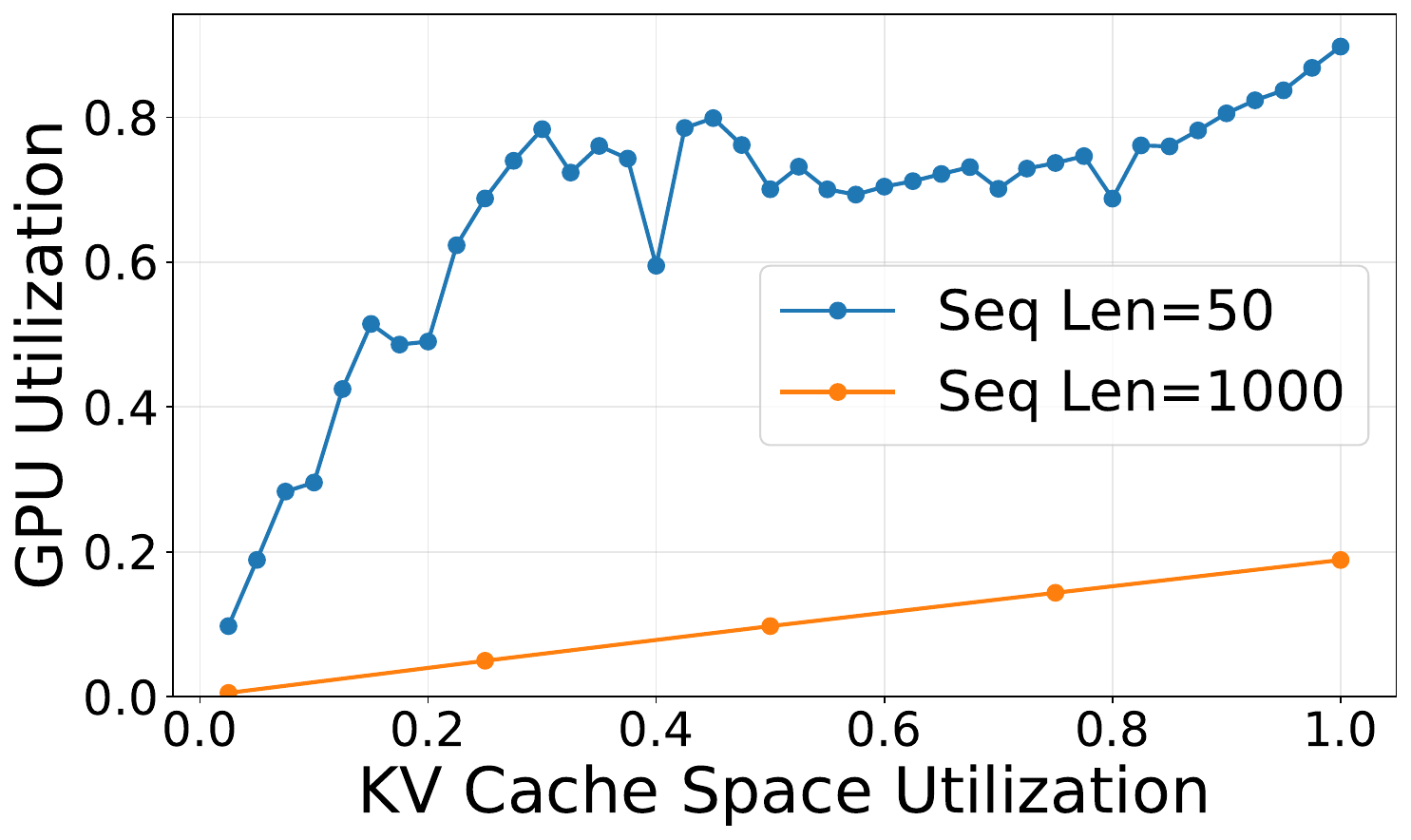}
        \label{fig:diverse_bounds}
    }
    \hfill
    \subfigure[Variation of the per-step attention-computation time
    as decoding proceeds.
    ]{
        \includegraphics[width=0.45\linewidth]{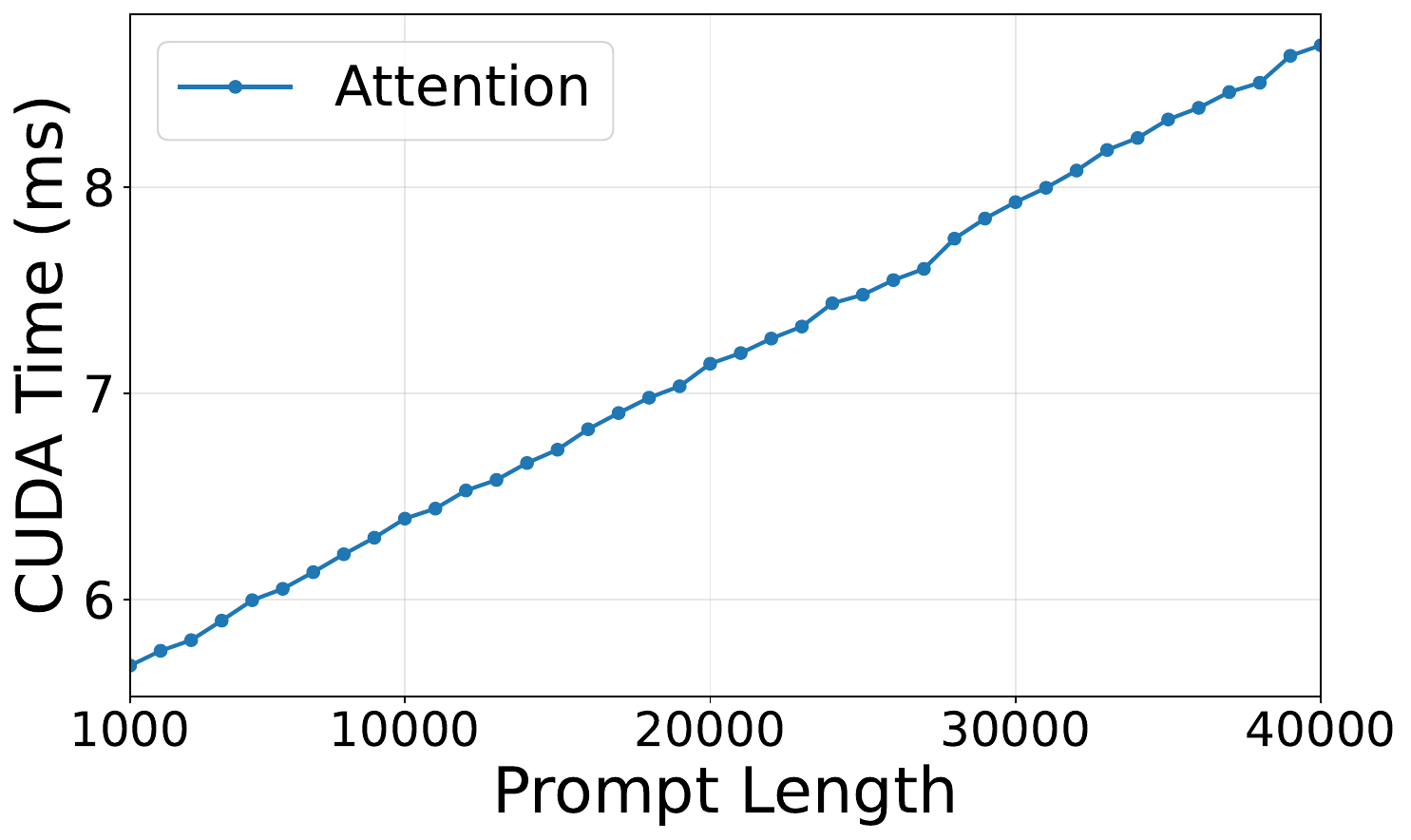}
        \label{fig:compute_cost}
    }
    \caption{Measurements on the instantaneous resource bound, as well as the compute cost characteristics in decoding. All measurements are conducted on an H800 GPU with Qwen3-32B model.} 
    \label{fig:bounds_and_observations}
\end{figure}

We note that for multi-resource cost modeling, it is crucial to first understand which resource is the true bottleneck, which determines the follow-up cost analysis road-map.  
In reality, an LLM service backend may be either \emph{compute-bound} or \emph{memory bound}---depending on the runtime workload characteristics.
Specifically, in Fig.~\ref{fig:diverse_bounds} we depict\footnote{
In Fig.~\ref{fig:bounds_and_observations}, GPU utilization is estimated by comparing the achieved TFLOPs to the peak TFLOPs, and the CUDA time is measured with \texttt{torch.profiler}.
}
the relationship between compute and memory consumptions for a decode step served atop an H100 GPU with increasing batch size.
We respectively fix the sequence length (including both the input length and the latest output length so far) to 50 and 1000.
Longer sequences tend to be more memory-intensive: when serving short sequences, the GPU utilization is already very high before reaching the memory limit; yet for long sequences, the GPU utilization has not yet ramped up when GPU memory is full.
Note that the service cost on a resource type matters only when that resource becomes the bottleneck. 
We therefore explore how to analyze the LLM service cost respectively in each status.

We first focus on inference cost modeling in \emph{memory-bound} status.
In such circumstances, the KVCache occupation of an LLM request is a key factor in cost modeling---less KVCache consumption allows more requests be concurrently served (with low GPU utilization, serving more requests concurrently won't increase the iteration time). 
Therefore, in memory-bound scenarios, we need to primarily consider a request's cumulative KVCache consumption during its inference process.
That is, let $I$ and $O$ be the input (i.e., prompt) and output token length of an inference, then its total service cost $\mathcal{C}$ can be depicted by $\mathcal{C}_M=\sum_{l=I}^{I+O}l*U_{MT}=(\frac{O^2}{2}+IO)U_{MT}$, where $U_{MT}$ represents the unit \emph{memory-time} product (i.e., the KVCache space occupied by a single token in a single step). 


We next turn to the \emph{compute-bound} status. 
In such cases, the KVCache occupation of an LLM inference is not a concern on the service backend, and we only need to quantify its total computation amount over its lifetime.
To that end, in Fig.~\ref{fig:compute_cost}, we measure the per-iteration attention-computation time (a key computing module in Transformers) in different decoding\footnote{
In this paper we focus on resource consumption in \emph{decoding} steps; with the wide adoption of \emph{reasoning LLMs}~\cite{cai2025r}, the decoding phase of an inference often dominates its TTLT.
Moreover, apart from the linear attention-computation time, there also exists a \emph{constant} FFN computation time (as well as some minor time factors like CUDA-lunching delay) that is irrelevant to the sequence length. 
However, the FFN time can be remarkably amortized in long decodes or with large batch sizes (given the GPU GEMM optimizations); so for simplicity we skip it here.  
} steps when an inference is served alone. 
It confirms that the duration is linear to the accumulated sequence length so far.
{Therefore, we can depict the total computation cost of an inference as $\mathcal{C}_C=\sum_{l=I}^{I+O}l*U_{CT}=(\frac{O^2}{2}+IO)U_{CT}$, where $U_{CT}$ represents the unit \emph{compute-time} product} (i.e., the average computation amount required to process a single KVCache token in a single step). 

Given the above analysis, we find that both the compute and memory costs share the same paradigm (although with different units, which does not affect the relative cost order among requests).
Therefore, we can adopt a uniform cost modeling method in practice (with no need to switch on the bottleneck pattern): $\mathcal{C}=\frac{O^2}{2}+IO$.
Despite such cross-resource paradigm consistency, we note that our formulation is substantially different from existing ones---e.g., purely $O$ in \cite{qiu2024efficient,fu2024efficient,shahout2024don}, or a simple weighted summation of $I$ and $O$ in \cite{sheng2024fairness}. 
Our later evaluations in {Sec.~\ref{sec:cost modeling}} confirm the superiority of such a modeling approach.

\subsection{Uncertainty-aware Request Scheduling}
\label{subsec:gittins}


With the above output-prediction and cost-modeling methods, we can now depict the cost-distribution of each LLM request upon its arrival (which can mitigate---yet not eliminate---demand uncertainty). 
Here we need to design an uncertainty-aware scheduling algorithm that can attain the theoretically optimal TTLT performance.

\begin{figure}[t]
    \centering
        \includegraphics[width=0.45\linewidth]{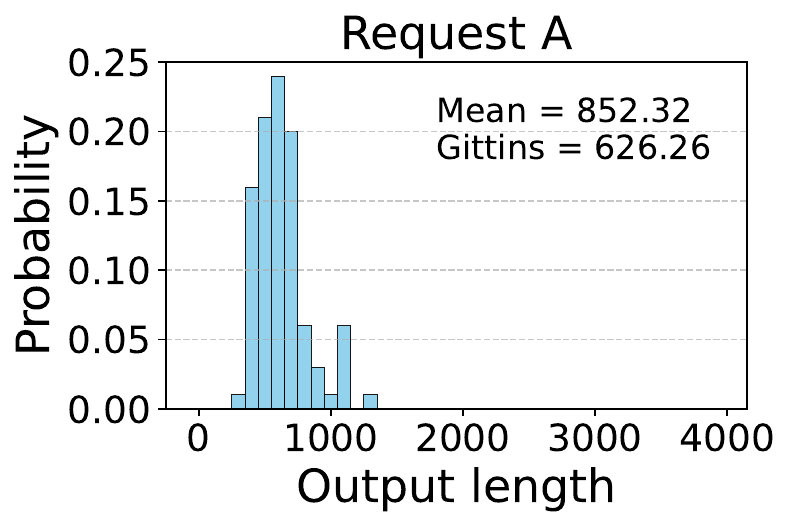}
        \includegraphics[width=0.45\linewidth]{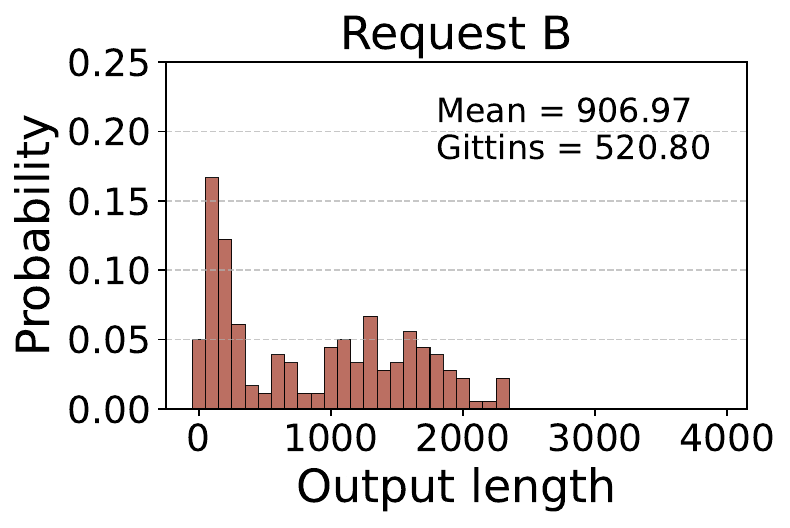}
    \caption{An example illustrating the deficiency of Mean-value-based request prioritization. While request-B has a larger Mean value, due to its irregular demand-distribution shape, however, serving it first can be expected to behave better in average TTLT.} 
    \label{fig:gittins}
\end{figure}

In request queuing, we need to calculate a proper scheduling index out of each request's cost distribution (for which the mean value is not optimal, as suggested in Fig.~\ref{fig:gittins}).
In fact,  we notice that selecting a highest-priority inference to serve is similar to selecting the most-rewarding arm to take in a \emph{multi-armed bandit problem}~\cite{mahajan2008multi}: the cost distribution is essentially a reward distribution depicting how allocating additional resources to a request can help to yield inference completion. 
For such a multi-armed bandit problem with nondeterministic reward but deterministic reward distribution, in the literature it is known that the \emph{Gittins policy}~\cite{gittins1979bandit} can achieve the theoretically optimal performance.

Specifically, given a request with a cost distribution of $\mathcal{D}$, the Gittins policy calculates its Gittins index defined as $G(\mathcal{D})=\operatorname*{inf}_{\Delta>0,X\sim\mathcal{D}}{\frac{\operatorname{E}[\operatorname*{min}\{X,\Delta\}]}{\operatorname*{P}(X\leq\Delta)}}$; requests with smaller Gittins indices are prioritized in scheduling. 
In that formula, $\operatorname{E}[\operatorname*{min}\{X,\Delta\}]$ represents the request service consumption when allocated a service budget $\Delta$, and $\operatorname*{P}(X\leq\Delta)$ represents its probability of successful completion before $\Delta$ is used up; $G$ thus represents the minimum amortized cost ever attainable for that request. 
It has been proven that, scheduling jobs based on such Gittins index (instead of the distribution expectation value) can yield the minimum average latency (TTLT)~\cite{gittins1989multiprocessor}. {The effect can be observed in Fig.~\ref{fig:gittins}, where the Gittins policy prioritizes the request that are more likely to complete in the near future.}

Furthermore, the Gittins index of each pending request shall be properly updated at runtime after inference commencement.
As indicated by TRAIL~\cite{shahout2024don} and FastServe~\cite{wu2023fast}, enabling preemption is necessary to optimize the average TTLT (with negligible overheads given recent techniques like \emph{swap-compute overlapping} and \emph{KVCache prefetching}).  
Ideally, we need to keep refreshing the Gittins index of each active inference at runtime---with the information of \emph{remaining} service cost distribution.
Yet, continuously doing so after each inference step is still impractical---due to the high monitoring overhead as well as the thrashing risk (i.e., the relatively scheduling order of two requests with similar Gittins index may frequently reverse).
We therefore divide each request's cost range into multiple (defaulted to 10) buckets; the Gittins index of each request is refreshed only at bucket boundaries. 
In this way, we can seek a proper balance between rescheduling timeliness and system stability.

%% file: 4-evaluation.tex
\section{Evaluation}
\label{sec:eval}

\subsection{Setup}
\phm{Hardware platform.}
In our testbed experiments, we respectively use Llama3.1-8B~\cite{llama3_1_8b_instruct} and Qwen3-32B~\cite{qwen3_32b} as the LLM backend; the Llama3.1-8B model is deployed on a server equipped with A40-PCIe-48GB GPU, and the Qwen3-32B model is deployed on a server with H800-PCIe-96GB GPU. 
Moreover, we also build a simulator of {up to 64 GPUs} to evaluate the scalability performance of \oursched.

\phm{Workloads.}
In our experiments we use three typical inference datasets as in \cite{hu2024inference}: SharedGPT~\cite{sharegpt_vicuna_unfiltered_2025}, Alpaca-Summarization~\cite{alpaca_pubmed_summarization_2025}, and Document-Write~\cite{write_doc_sft_v1_2025}. 
The characteristics of those datasets are previously elaborated in Fig.~\ref{fig:background_hybridity}. 
Regarding request submission interval, we employ a Poisson distribution under different $\lambda$ hyper-parameters (yielding diverse task rates).

\phm{Baselines.}
We compare our \oursched solution with multiple representative baselines: FCFS~\cite{kwon2023efficient}, FastServe~\cite{wu2023fast}, SSJF~\cite{qiu2024efficient}, LTR~\cite{fu2024efficient}, and TRIAL~\cite{shahout2024don}.
All of those baselines have been elaborated in Sec.~\ref{subsec:limitations}, and their hyper-parameters are all set as default. 
Besides, by default, for \oursched the selection ratio in predictor is set to 0.8, and the bucket size for Gittins analysis is set to 200.

\phm{Metrics.}
As elaborated in Sec.~\ref{subsec:problem}, we use the average TTLT 
as the primary scheduling metric.
That said, we also evaluate the average TTFT 
performance of each scheduler.

\subsection{End-to-end Experiment}

\begin{figure}[t]
    \centering
    
    \subfigure[Llama3.1-8B]{
        \includegraphics[width=0.45\linewidth]{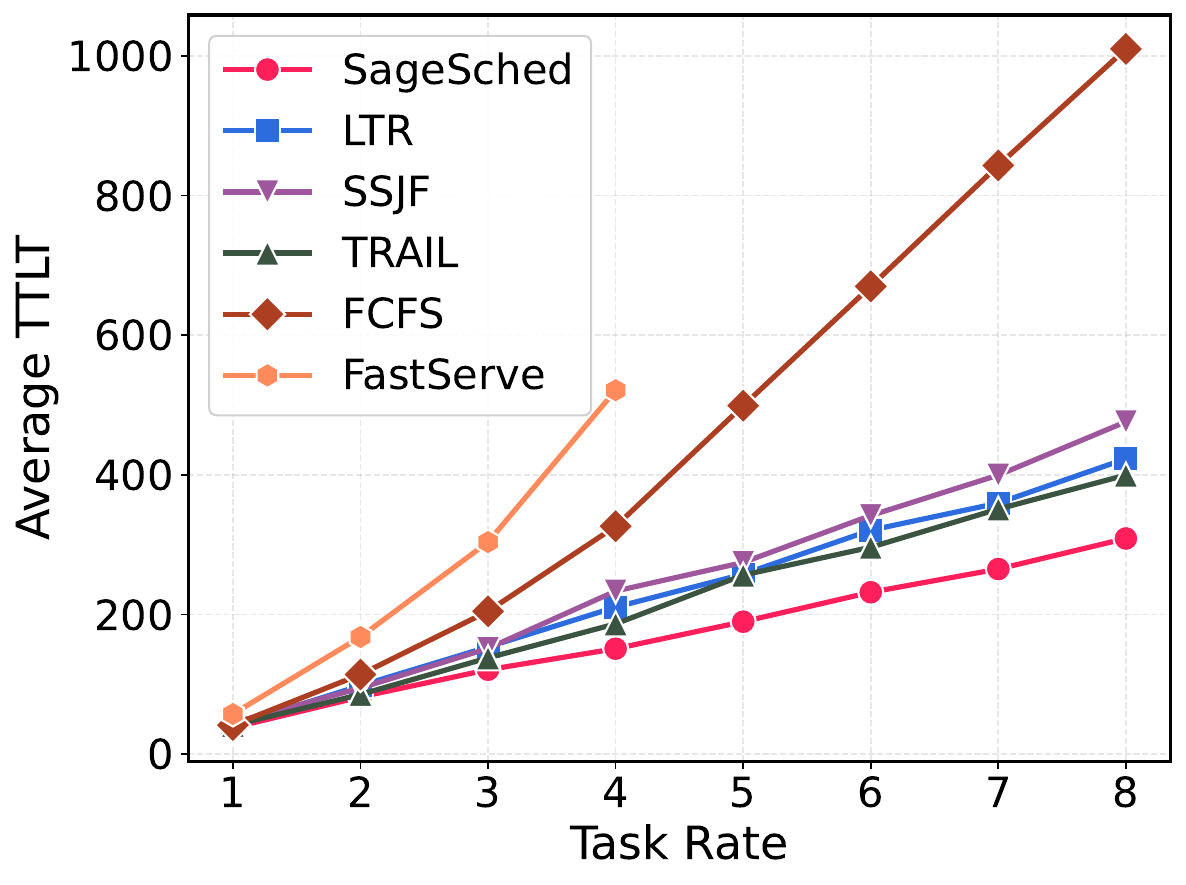}
        \includegraphics[width=0.45\linewidth]{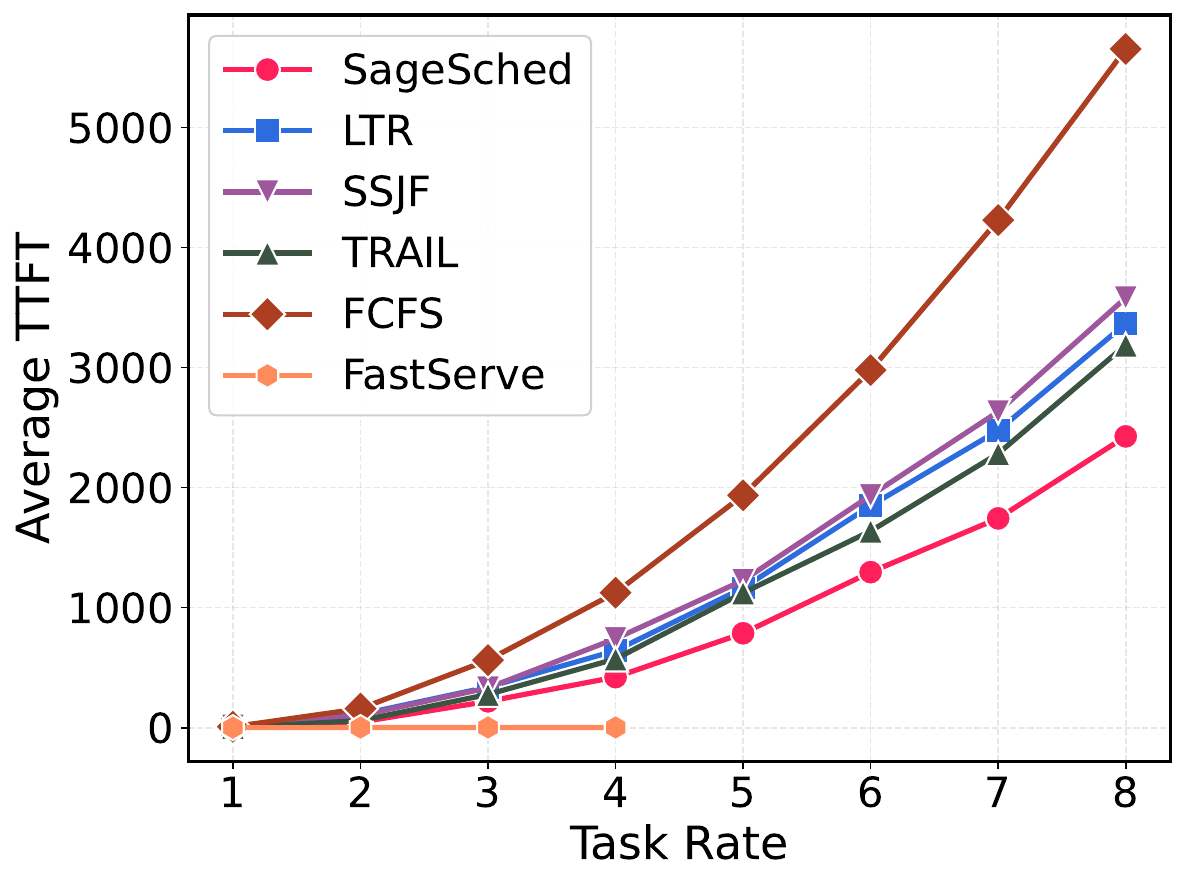}
        \label{fig:e2e 8b ttft}
    }

    \subfigure[Qwen3-32B]{
        \includegraphics[width=0.45\linewidth]{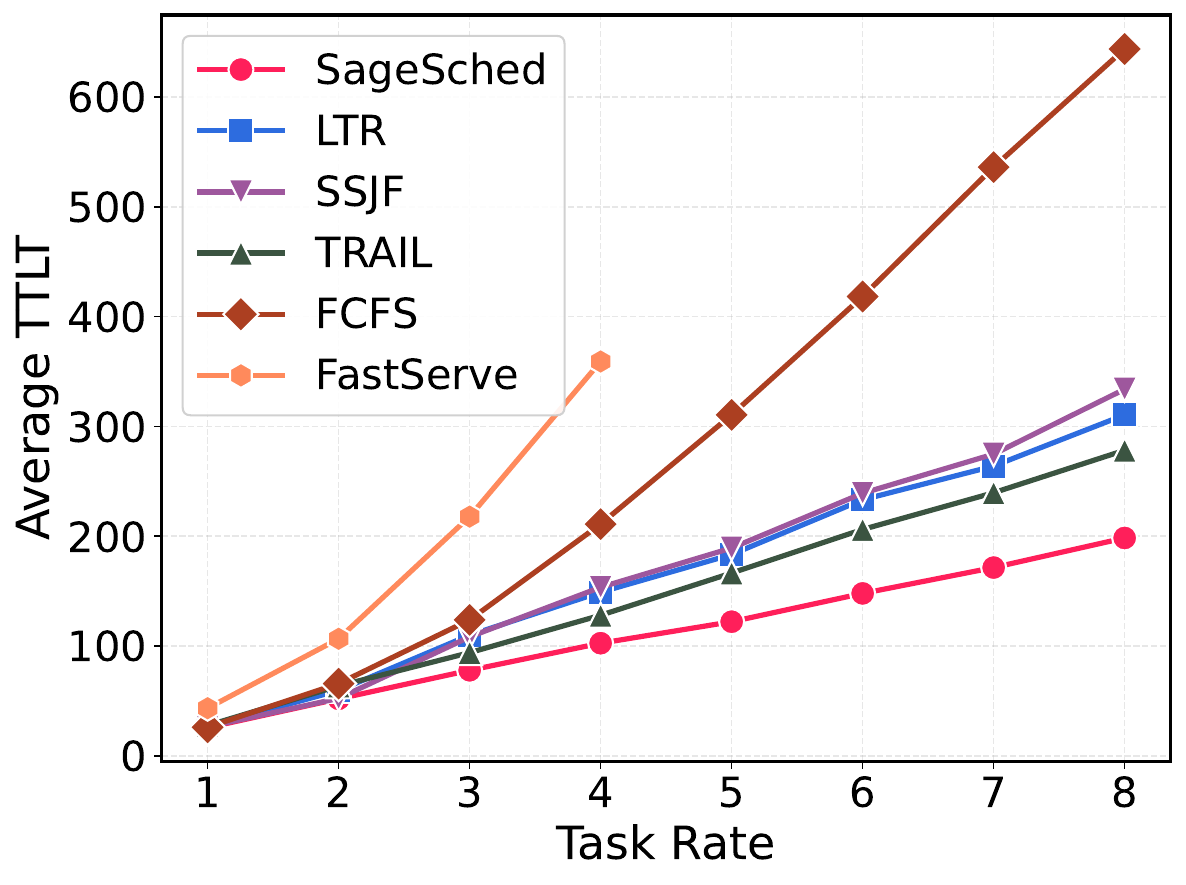}
        \includegraphics[width=0.45\linewidth]{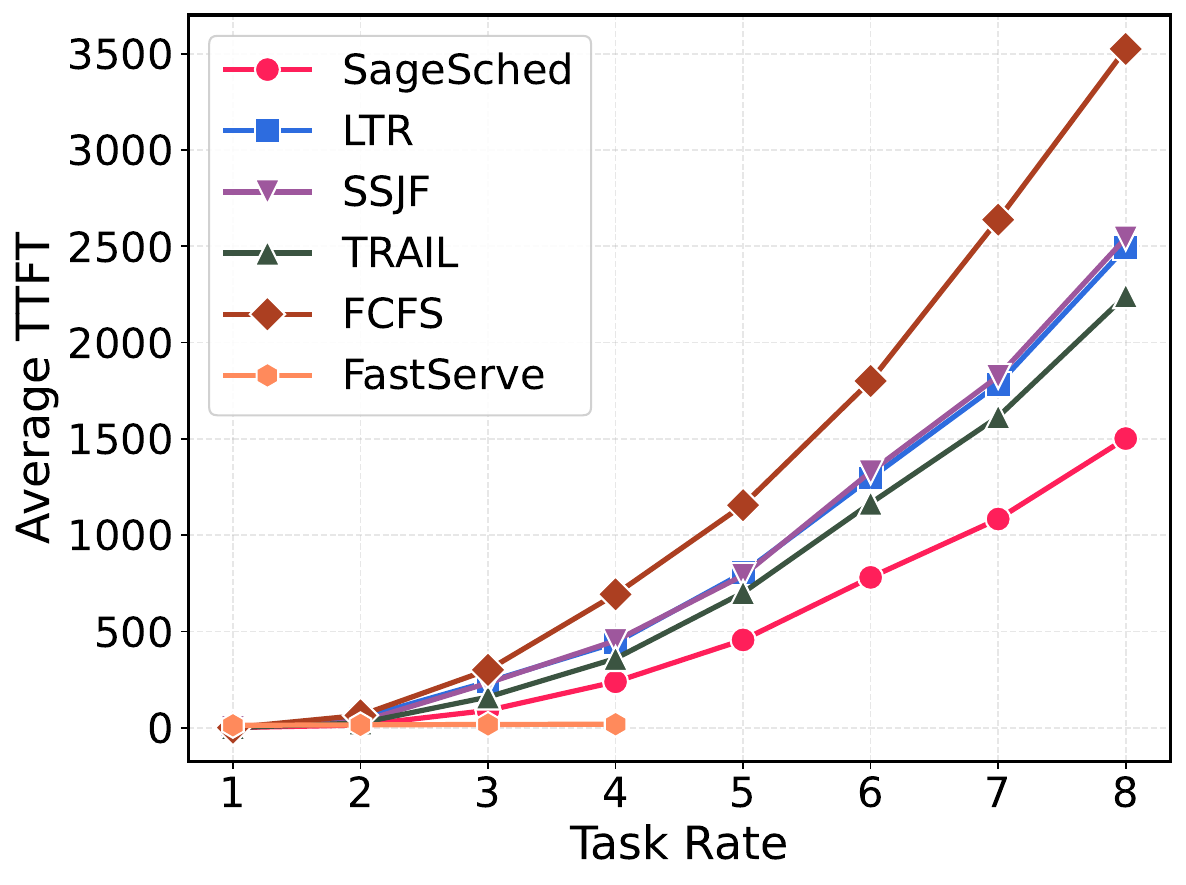}
        \label{fig:e2e 32b ttft}
    }

    \caption{End-to-end performance on mixed datasets.}
    \label{fig:e2e}
\end{figure}

\phm{Experiments with requests from mixed datasets.}
In our end-to-end evaluation, we first merge the three datasets together, the requests from which are then randomly submitted; we experiment respectively
under different task rates. 
As shown in Fig.~\ref{fig:e2e}, \oursched demonstrates the best scheduling efficiency in TTLT across all the setups.
For example, when inferences are submitted at a rate of 8 requests-per-second (RPS) to the Qwen3-32B model, \oursched surpasses the second best scheduler, TRAIL, by 28.7\%. 
Such improvements are higher in cases with more intensive competitions, because \oursched can consistently work out the best scheduling order by properly handling the uncertainty and hybridity characteristics of LLM inference demands.

Regarding TTFT, while it is not the primary focus of this paper, Fig.~\ref{fig:e2e} indicates that \oursched can also attain a satisfying performance, because it can effectively mitigate the head-of-line-blocking problem. 
In particular, although FastServe---by always prioritizing the newly-arrival requests---can attain the best TTFT, in TTLT, it however behaves much worse than other schedulers due to its interleaved execution mode with MLFQ. 
In that sense, \oursched is the only scheduler that can behave well in both TTLT and TTFT.

\begin{figure}
    \centering
    \includegraphics[width=1\linewidth]{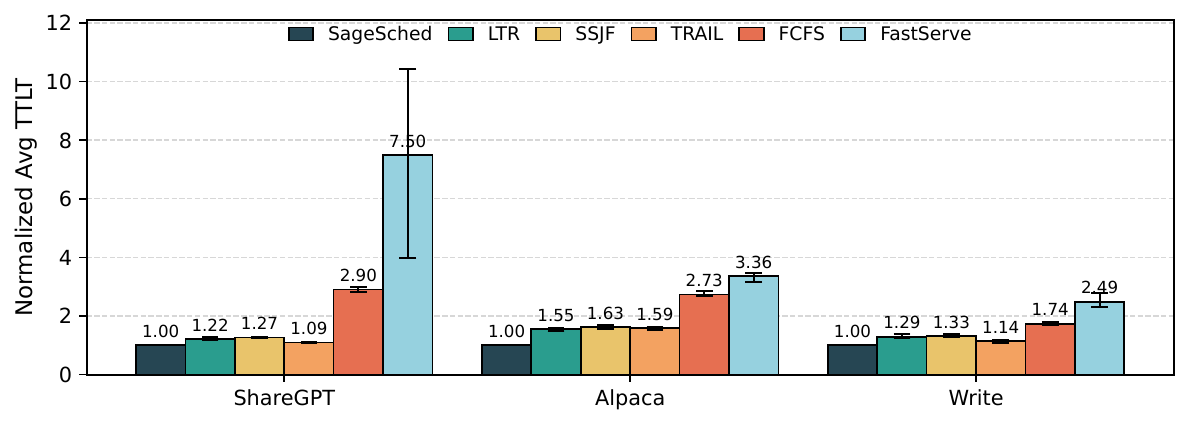}
    \caption{End-to-end performance respectively with each dataset.} 
    \label{fig:single_dataset}
\end{figure}

\phm{Experiments with request respectively from each dataset.}
Recall that the three datasets we use have different input-output length distribution (Fig.~\ref{fig:background_hybridity}); we thus experiment respectively with each dataset, with the results shown in Fig.~\ref{fig:single_dataset}.
We first note that \oursched consistently perform the best for each dataset.
In particular, its performance superiority peaks for the Alpaca (Summarization) dataset: this is because requests in Alpaca dataset typically have long input lengths, for which existing output-length based scheduling methods are highly inaccurate in cost modeling.

\subsection{Microscopic Deep Dive}

In this part, we dive deep into the effectiveness of each solution component in \oursched.
Unless otherwise specified, we use the same setup as in Fig.~\ref{fig:e2e} (under a RPS of 8).

\subsubsection{Superiority of our Predictor Design}

\begin{figure}
    \centering
    \begin{minipage}[t]{0.47\linewidth}
        \centering
        \includegraphics[width=\linewidth]{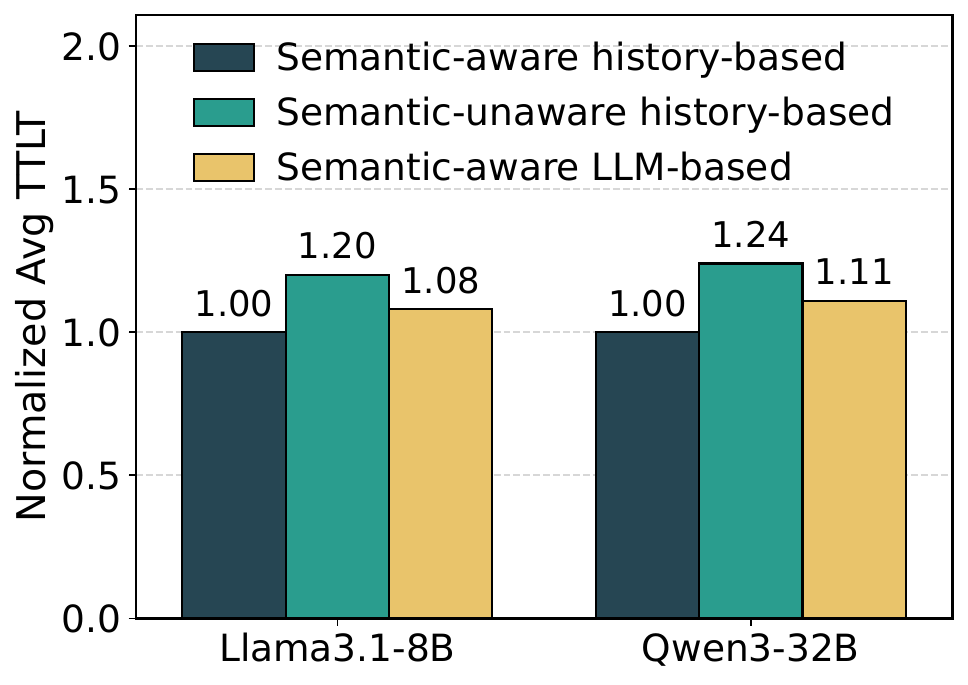}
        \caption{Performance comparison of different output-length prediction methods.} 
        \label{fig:ablation_prediction}
    \end{minipage}
    \hfill
    \begin{minipage}[t]{0.47\linewidth}
         \centering
        \includegraphics[width=\linewidth]{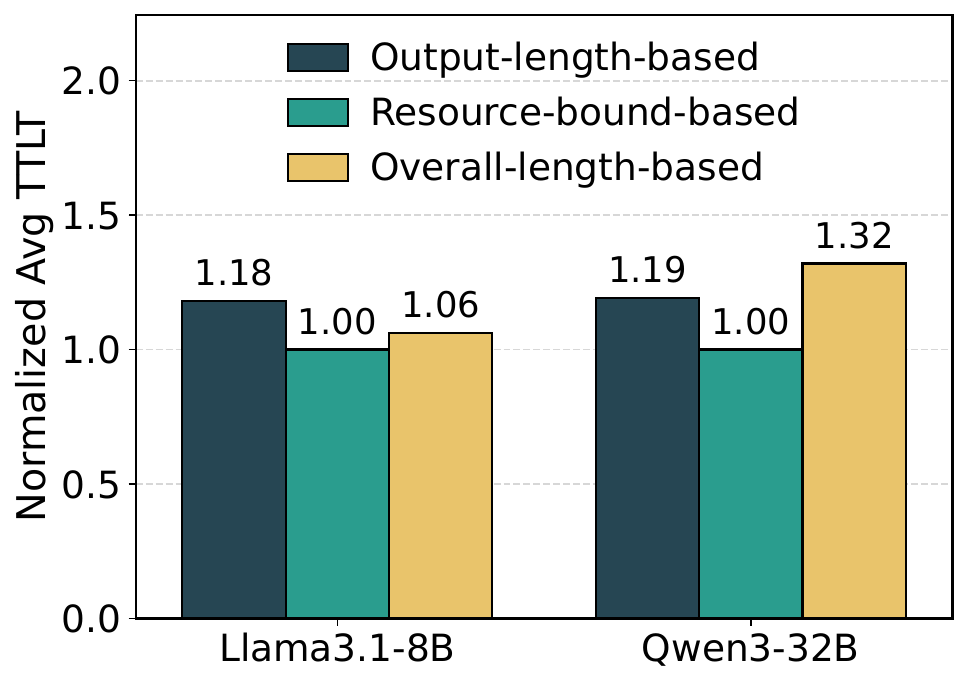}
        \caption{Performance comparison of different service cost modeling methods.}
        \label{fig:ablation_cost}
    \end{minipage}
    \hfill
\end{figure}

Given the deficiency of conventional output-length predictors, in Sec.~\ref{subsec:predictor} we propose a novel \emph{semantic-aware history-based predictor}.
Here we check its necessity with regard to the end-to-end TTLT performance.
Specifically, we incorporate two baselines representing the other prediction choices: (1) \emph{semantic-unaware history-based predictor}---which predicts the output-length length of a request by referring to the historical requests with a similar input length (following the same filtering method as us), and (2) \emph{semantic-aware LLM-based predictor}---still using a DistillBert model as in \cite{qiu2024efficient} yet, to predict a distribution, we remove the \texttt{logits.argmax} layer.
In Fig.~\ref{fig:ablation_prediction}, we show the performance of \oursched respectively with different predictors.
It shows that with our semantic-aware history-based predictor, due to its high distribution-prediction accuracy, \oursched can make the best TTLT performance.

Moreover, we note that our predictor also surpasses others in training and prediction overheads.
First, our method completely eliminates training overhead, whereas for the LLM-based predictor, it takes several hours to construct a training dataset and also tens of minutes to perform model fine-tuning.
Second, the average per-request prediction latency is below 0.5 ms under our predictor (consisting of 0.22 ms for semantic embedding extraction and 0.15 ms for vector retrieval), yet for the LLM-based predictor it is approximately 3.6 ms.
Those results above jointly confirm that our semantic-aware history-based predictor is light-weight and also accurate.

\subsubsection{Superiority of our Cost Modeling Method}
\label{sec:cost modeling}

In Sec.~\ref{subsec:cost}, after analyzing both the compute- and memory-bound cases,
we developed a \emph{resource-bound-based modeling} method to depict LLM inference cost.
To confirm its necessity, we replace it respectively to two other methods: (1) \emph{output-length-based modeling}---which directly uses the output length as the inference cost~\cite{shahout2024don,qiu2024efficient}, and (2) \emph{overall-length-based modeling}, which combines the input and output lengths as the cost (we assign the output length a doubled weight as in \cite{sheng2024fairness}). 
Fig.~\ref{fig:ablation_cost} shows the overall TTLT performance under different modeling methods. 
It confirms the necessity to adopt the resource-bound-based method for cost modeling.

\subsubsection{Superiority of our Scheduling Policy}

\begin{figure}[t]
    \centering
    \begin{minipage}[t]{0.47\linewidth}
        \centering
        \includegraphics[width=\linewidth]{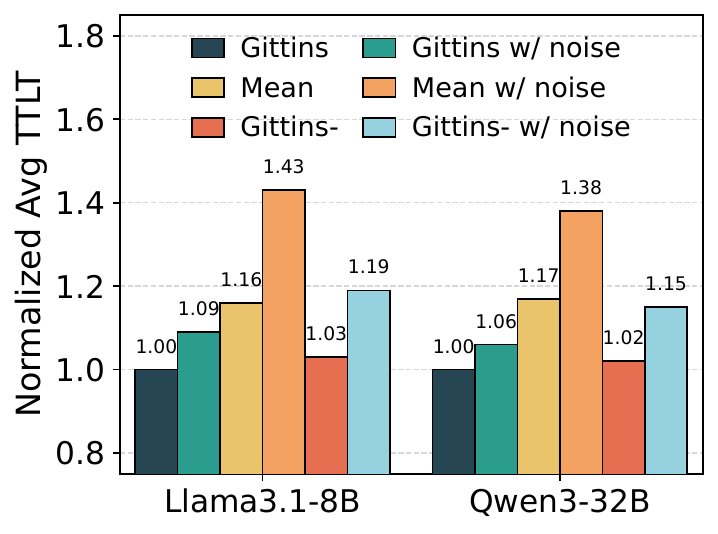 }
        \caption{Comparison of different scheduling methods, even with added cost noises.} 
        \label{fig:ablation_gittins}
    \end{minipage}
    \hfill
    \begin{minipage}[t]{0.47\linewidth}
        \centering
        \includegraphics[width=\linewidth]{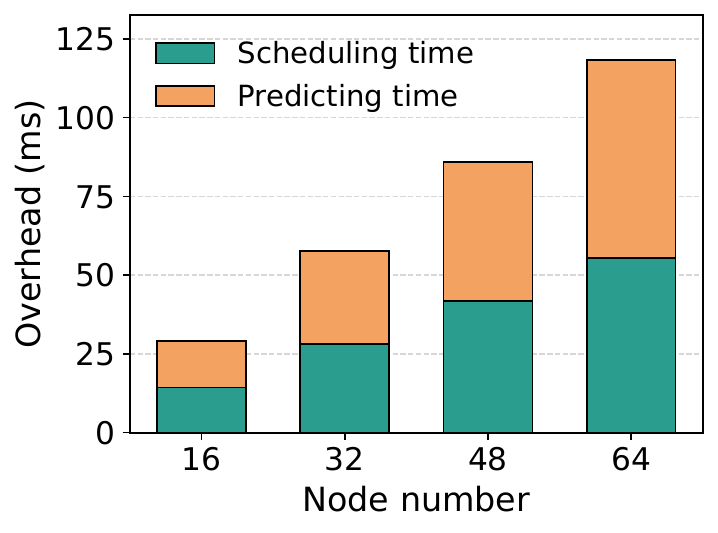}
        \caption{Predicting and scheduling overheads of \oursched at different scales.} 
        \label{fig:overhead}
    \end{minipage}
    
\end{figure}

Recall that our uncertainty-aware scheduling algorithm in Sec.~\ref{subsec:gittins} has two key techniques: 
\emph{Gittins-index-based ordering}, and \emph{runtime Gittins-index refreshing}.
Here we respectively evaluate the benefit of each technique; besides, we also need to check the robustness of scheduling quality in cases with inaccurate predictions.
Specifically, we first introduce two additional baselines: (1) \emph{Mean}---which orders requests based on the \emph{mean} value of their cost distributions, and (2) \emph{Gittins}$-$---which applies the Gittins policy but without timely refreshing the Gittins index at bucket boundaries.
Moreover, for each scheduler, we also measure its performance against \emph{inaccurately-predicted} cost distributions (by merging a \emph{uniform} distribution to the original distribution following a weight ratio of 1:4). 
As shown in Fig.~\ref{fig:ablation_gittins}, both Gittins-index-based ordering and runtime Gittins-index refreshing are beneficial for TTLT performance; besides, the added noises incur much less performance degradation to our Gittins-index-based algorithm than to other baselines.

\subsection{Overhead and Sensitivity Analysis}
\label{subsec:eval_sensitivity}

In this part we study the scalability issue of \oursched and also make sensitivity analysis on the hyper-parameters introduced. 
Unless otherwise specified, we also use the same setup as in Fig.~\ref{fig:e2e}. 

\phm{Scalability performance.}
To evaluate the scalability performance of \oursched, we simulate large-scale cluster deployments with up to 64 GPU nodes. 
The system load (RPS) is increased proportionally with the cluster size.
On average, each node serves 8 requests per second (RPS), with up to 1,000 requests buffered in the queue.
We measure the per-request latency incurred during the predicting and scheduling stages, assuming a fixed output length of 1,000 tokens. 
As shown in Fig.~\ref{fig:overhead}, the latency grows only linearly as the cluster scales. 
Even at the largest scale of 64 nodes, the average additional latency per request remains around 100 ms.
Given that typical LLM inference workloads operate at the granularity of several seconds to minutes (as in Fig.~\ref{fig:e2e}), this overhead is indeed negligible.
Moreover, for large clusters, we can deploy multiple concurrent schedulers to further alleviate such overhead.

\begin{figure}[t]
    \centering
    \subfigure[Predictor hyperparameter]{
        \includegraphics[width=0.45\linewidth]{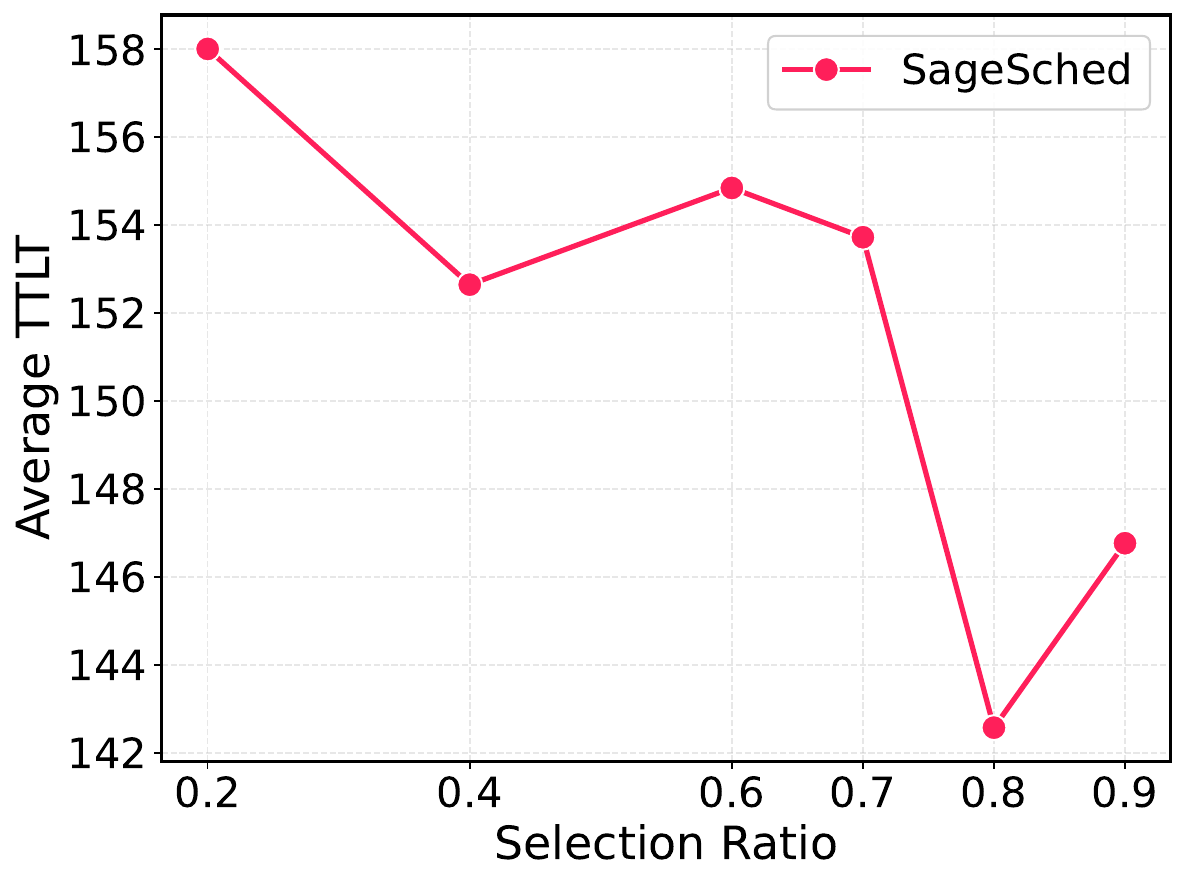}
        \label{fig:sensitivity_predictor}
    }
    \hfill
    \subfigure[Scheduler hyperparameter]{
        \includegraphics[width=0.45\linewidth]{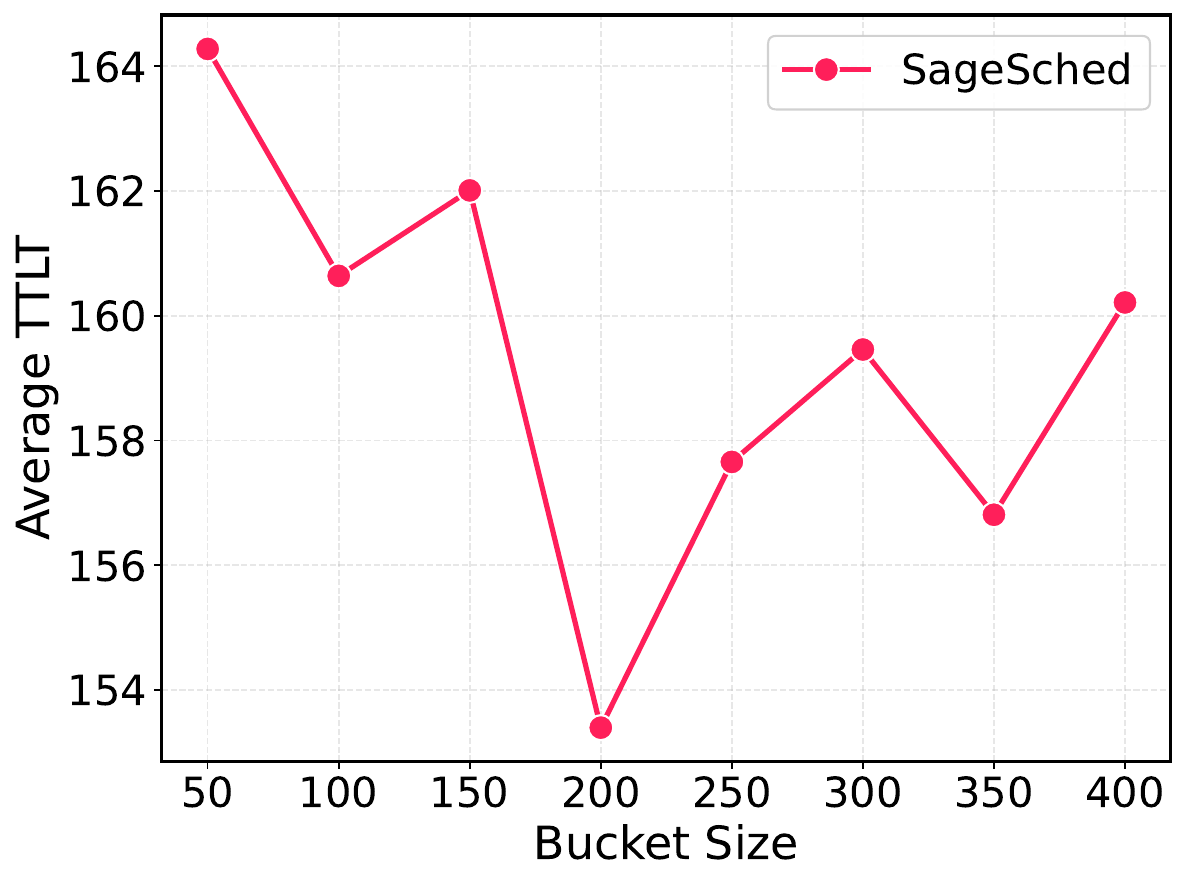}
        \label{fig:sensitivity_gittins}
    }
    \vspace{-.1in}
    \caption{Sensitivity analysis of \oursched hyperparameters.}
    \label{fig:sensitivity}
\end{figure}

\phm{Similarity threshold.} 
In predictor design (Sec.~\ref{subsec:predictor}) we adopt a cosine-similarity threshold in filtering out the historical requests with similar prompts. 
Here we respectively set it to different values, with the testbed performance shown in Fig.~\ref{fig:sensitivity_predictor}.
The results confirm that setting it to 0.8 yield the best performance---an over-large threshold renders the selected requests less abundant (compromising the capturing of demand uncertainty), yet an over-small threshold would pollute the sampled distribution with irrelevant requests.

\phm{Bucket size.}
Meanwhile, in Sec.~\ref{subsec:gittins} we also adopt a bucket-size hyper-parameter to control the refreshing frequency of Gittins index. 
As in Fig.~\ref{fig:sensitivity_gittins}, we measure the average TTLT under \oursched with different bucket sizes.
It suggests that a bucket size of 200 can yield the best performance (which we thus adopt as default): smaller bucket sizes lead to more priority calculation and re-scheduling overheads, yet larger bucket sizes compromise the accuracy of priority assignment; both degrade the overall efficiency.

%% file: 5-related.tex
\section{Additional Related Works}
\label{sec:related}

A broad range of studies aim to improve the efficiency of LLM serving from complementary angles; they can be combined with \oursched.
For example, some works seek to accelerate individual inferences from per-token level~\cite{dao2022flashattention,dao2023flashattention,dettmers2022gpt3,frantar2022gptq,leviathan2023speculative,miao2023specinfer} or with deployment optimizations~\cite{zhong2024distserve,qin2024mooncake,hu2024memserve,hu2024inference,agrawal2024taming}.
Some others seek to improve backend utilization by redesigning batching and admission control mechanisms~\cite{duan2024muxserve,xiang2025aegaeon}.
Besides, since KVCache storage is often a dominant bottleneck, many techniques have also been proposed to reduce KVCache \emph{size} via compression or selective retention~\cite{xiao2023efficient,zhang2023h2o,lee2024infinigen}.
Due to the space limitation, we place the detailed elaboration of those related works to appendix.

%% file: 6-conclusion.tex
\section{Conclusion}
\label{sec:conclusion}
In this work, we present \oursched, an efficient scheduler tailored for the uncertainty and hybridity characteristics of LLM workloads.
\oursched introduces a semantic-aware history-based predictor to predict output-length distribution in a light-weight and also accurate manner.
It also models the overall service cost of LLM inferences with both memory and compute bounds considered.
Besides, \oursched adopts an uncertainty-aware scheduling policy which leverages the Gittins index for the best queueing performance.
Testbed and simulation experiments confirm that \oursched can substantially improve the end-to-end efficiency performance, with an TTLT reduction of over 28.7\%.


%% file: 7-appendix.tex
\newpage
\appendix
\onecolumn

\section{A Full Version of Related Works}
Our work focuses on \emph{request-level scheduling} under uncertain and hybrid (compute--memory) demands; the following lines of research are largely orthogonal and can be combined with \oursched.

\phm{LLM inference acceleration.}
A substantial body of work reduces the \emph{per-token} computation and memory cost of LLM inference. At the kernel level, FlashAttention and FlashAttention-2 reorganize attention computation to improve memory locality and reduce redundant reads/writes, yielding faster and more memory-efficient exact attention~\cite{dao2022flashattention,dao2023flashattention}. Model compression and low-bit quantization further reduce memory footprint while retaining accuracy, e.g., INT8 matrix multiplication and post-training quantization for transformer weights~\cite{dettmers2022gpt3,frantar2022gptq}. Recent decoding-side acceleration methods exploit \emph{speculative execution}: speculative decoding uses a fast draft model to propose multiple tokens and then verifies them with the target model~\cite{leviathan2023speculative}, while subsequent systems improve speculative serving efficiency via token-tree verification and online mechanisms~\cite{miao2023specinfer,liu2023online}. 

Beyond token-level acceleration, system designs optimize the end-to-end pipeline. Prefill--decode disaggregation decouples throughput-oriented prefill from latency-critical decoding and restructures KVCache placement across devices~\cite{zhong2024distserve,qin2024mooncake,hu2024memserve,hu2024inference}, while chunked-prefill overlaps long-context prefill with decoding and improves cache reuse to better manage throughput--latency trade-offs~\cite{agrawal2024taming,hu2024inference}. 

These techniques improve the \emph{execution efficiency} of each request; \oursched is complementary by deciding \emph{which} request to serve next given bounded resources and variable request lengths.

\phm{LLM serving architectures and multiplexing.}
Many systems improve backend utilization by redesigning batching, admission control, and multi-tenant resource sharing. Foundational serving frameworks (e.g., Orca-style continuous batching and modern engines such as vLLM and SGLang) emphasize high device utilization and efficient KVCache management~\cite{yu2022orca,kwon2023efficient,zheng2024sglang}. For multi-model and multi-tenant settings, recent work explores finer-grained multiplexing and pooling: MuxServe enables flexible spatial--temporal multiplexing to co-locate multiple LLM workloads on shared GPUs~\cite{duan2024muxserve}, and Aegaeon proposes effective GPU pooling for concurrent LLM serving on the market~\cite{xiang2025aegaeon}. 

These systems primarily address \emph{resource sharing and placement} across models and tenants; \oursched instead targets \emph{within-backend} request ordering under uncertain service times, and can be integrated as the per-GPU (or per-pool) scheduler.

\phm{LLM request scheduling and fairness.}
A growing literature studies request scheduling to mitigate head-of-line blocking and improve tail latency. Production engines often start with FCFS, which is simple but can severely degrade TTLT under heterogeneous request lengths~\cite{kwon2023efficient,zheng2024sglang}. FastServe approximates SRPT using an MLFQ-like policy with periodic priority demotion, trading latency gains for additional preemption overheads~\cite{wu2023fast}. More recent schedulers adopt \emph{predict-and-order}: $S^3$ uses a finetuned small model to estimate generation length~\cite{jin2023s}, PO queries an LLM to anticipate its output length~\cite{po}, SSJF predicts output length with a proxy model to approximate SJF~\cite{qiu2024efficient}, and learning-to-rank approaches predict relative length ranks rather than exact lengths~\cite{fu2024efficient}. Embedding-based schedulers further exploit prompt features extracted from LLM layers to improve prioritization~\cite{shahout2024don}. 
In parallel, fairness-aware serving formalizes the tension between efficiency and service differentiation and proposes scheduling mechanisms to meet fairness objectives~\cite{sheng2024fairness}. Llumnix addresses dynamic scheduling by incorporating runtime conditions and enabling adaptive reordering for better service quality~\cite{sun2024llumnix}. 

Compared to these works, \oursched differs in three aspects: (i) it targets \emph{distributional} (rather than point) perception of output length, avoiding fragile single-value predictions; (ii) it models \emph{hybrid} compute--memory costs induced by KVCache growth, rather than using output length alone as the service cost; and (iii) it explicitly incorporates IO impact into preemption decisions, aiming to avoid excessive swapping while still correcting priority.

\phm{KVCache management and long-context serving.}
Since decoding repeatedly attends to all previous tokens, KVCache is often the dominant memory bottleneck. PagedAttention (as implemented in vLLM) improves KVCache allocation by paging KV blocks to reduce fragmentation and improve memory efficiency~\cite{kwon2023efficient}. Beyond allocation, systems reduce KVCache \emph{size} via compression or selective retention: attention sinks enable efficient streaming by stabilizing attention patterns~\cite{xiao2023efficient}, and heavy-hitter based methods retain the most important tokens to approximate full attention under memory constraints~\cite{zhang2023h2o}. More recent work dynamically manages KVCache across time: InfiniGen proposes dynamic KVCache management for efficient generative inference~\cite{lee2024infinigen}, while KIVI quantizes KV cache asymmetrically to 2-bit without tuning, reducing memory pressure and bandwidth~\cite{liu2024kivi}. 

These advances improve the feasible concurrency and per-step efficiency; \oursched is orthogonal and can further improve TTLT by prioritizing requests using cost distributions that implicitly reflect KVCache growth and resource contention.

%% file: main.bib
@article{hu2024inference,
  title={Inference without interference: Disaggregate llm inference for mixed downstream workloads},
  author={Hu, Cunchen and Huang, Heyang and Xu, Liangliang and Chen, Xusheng and Xu, Jiang and Chen, Shuang and Feng, Hao and Wang, Chenxi and Wang, Sa and Bao, Yungang and others},
  journal={arXiv preprint arXiv:2401.11181},
  year={2024}
}

@article{wu2023fast,
  title={Fast distributed inference serving for large language models},
  author={Wu, Bingyang and Zhong, Yinmin and Zhang, Zili and Liu, Shengyu and Liu, Fangyue and Sun, Yuanhang and Huang, Gang and Liu, Xuanzhe and Jin, Xin},
  journal={arXiv preprint arXiv:2305.05920},
  year={2023}
}

@article{fu2024efficient,
  title={Efficient llm scheduling by learning to rank},
  author={Fu, Yichao and Zhu, Siqi and Su, Runlong and Qiao, Aurick and Stoica, Ion and Zhang, Hao},
  journal={Advances in Neural Information Processing Systems},
  volume={37},
  pages={59006--59029},
  year={2024}
}

@article{chen2023typefly,
  title={Typefly: Flying drones with large language model},
  author={Chen, Guojun and Yu, Xiaojing and Ling, Neiwen and Zhong, Lin},
  journal={arXiv preprint arXiv:2312.14950},
  year={2023}
}

@inproceedings{wang2024llm,
  title={Llmˆ3: Large language model-based task and motion planning with motion failure reasoning},
  author={Wang, Shu and Han, Muzhi and Jiao, Ziyuan and Zhang, Zeyu and Wu, Ying Nian and Zhu, Song-Chun and Liu, Hangxin},
  booktitle={2024 IEEE/RSJ International Conference on Intelligent Robots and Systems (IROS)},
  pages={12086--12092},
  year={2024},
  organization={IEEE}
}

@inproceedings{kwon2023efficient,
  title={Efficient memory management for large language model serving with pagedattention},
  author={Kwon, Woosuk and Li, Zhuohan and Zhuang, Siyuan and Sheng, Ying and Zheng, Lianmin and Yu, Cody Hao and Gonzalez, Joseph and Zhang, Hao and Stoica, Ion},
  booktitle={Proceedings of the 29th symposium on operating systems principles},
  pages={611--626},
  year={2023}
}

@article{jin2023s,
  title={$S^3$: Increasing GPU Utilization during Generative Inference for Higher Throughput},
  author={Jin, Yunho and Wu, Chun-Feng and Brooks, David and Wei, Gu-Yeon},
  journal={Advances in Neural Information Processing Systems},
  volume={36},
  pages={18015--18027},
  year={2023}
}

@inproceedings{yu2022orca,
  title={Orca: A distributed serving system for $\{$Transformer-Based$\}$ generative models},
  author={Yu, Gyeong-In and Jeong, Joo Seong and Kim, Geon-Woo and Kim, Soojeong and Chun, Byung-Gon},
  booktitle={16th USENIX Symposium on Operating Systems Design and Implementation (OSDI 22)},
  pages={521--538},
  year={2022}
}

@inproceedings{shahout2024don,
  title={Don't Stop Me Now: Embedding Based Scheduling for LLMs},
  author={Shahout, Rana and Malach, Eran and Liu, Chunwei and Jiang, Weifan and Yu, Minlan and Mitzenmacher, Michael},
  booktitle={ICLR},
  year={2025}
}

@INPROCEEDINGS{Recasens2025Mind,
  author={Recasens, Pol G. and Agullo, Ferran and Zhu, Yue and Wang, Chen and Lee, Eun Kyung and Tardieu, Olivier and Torres, Jordi and Berral, Josep Ll.},
  booktitle={2025 IEEE 18th International Conference on Cloud Computing (CLOUD)}, 
  title={Mind the Memory Gap: Unveiling GPU Bottlenecks in Large-Batch LLM Inference}, 
  year={2025},
  volume={},
  number={},
  pages={277-287},
  doi={10.1109/CLOUD67622.2025.00036}}

@misc{sharegpt_vicuna_unfiltered_2025,
  title        = {ShareGPT\_Vicuna\_unfiltered Dataset},
  author       = {{Hugging Face Datasets}},
  year         = {2025},
  howpublished = {\url{https://huggingface.co/datasets/anon8231489123/ShareGPT_Vicuna_unfiltered}},
  note         = {Accessed: 2026-01-27}
}

@misc{cpu_scheduling_comparison_gfg_2026,
  title        = {Comparison of Different CPU Scheduling Algorithms in Operating Systems},
  author       = {GeeksforGeeks},
  year         = {2026},
  howpublished = {\url{https://www.geeksforgeeks.org/comparison-of-different-cpu-scheduling-algorithms-in-os/}},
  note         = {Accessed: 2026-01-27}
}

@misc{alpaca_pubmed_summarization_2025,
  title        = {Alpaca-PubMed Summarization Dataset},
  author       = {{Hugging Face Datasets}},
  year         = {2025},
  howpublished = {\url{https://huggingface.co/datasets/ZhongshengWang/Alpaca-pubmed-summarization}},
  note         = {Accessed: 2026-01-27}
}

@misc{write_doc_sft_v1_2025,
  title        = {write\_doc\_sft\_v1 Dataset},
  author       = {{Hugging Face Datasets}},
  year         = {2025},
  howpublished = {\url{https://huggingface.co/datasets/lancexiao/write_doc_sft_v1}},
  note         = {Accessed: 2026-01-27}
}

@article{gittins1989multiprocessor,
  title={Multi-armed bandit allocation indices},
  author={Gittins, John C},
  journal={Wiley-Interscience Series in Systems and Optimization},
  year={1989},
  publisher={John Wiley \& Sons, Inc.}
}

@article{gittins1979bandit,
  title={Bandit processes and dynamic allocation indices},
  author={Gittins, John C and Jones, David M},
  journal={Journal of the Royal Statistical Society: Series B (Methodological)},
  volume={41},
  number={2},
  pages={148--164},
  year={1979},
  publisher={Wiley Online Library}
}

@inproceedings{sheng2024fairness,
  title={Fairness in serving large language models},
  author={Sheng, Ying and Cao, Shiyi and Li, Dacheng and Zhu, Banghua and Li, Zhuohan and Zhuo, Danyang and Gonzalez, Joseph E and Stoica, Ion},
  booktitle={18th USENIX Symposium on Operating Systems Design and Implementation (OSDI 24)},
  pages={965--988},
  year={2024}
}

@article{zheng2024sglang,
  title={Sglang: Efficient execution of structured language model programs},
  author={Zheng, Lianmin and Yin, Liangsheng and Xie, Zhiqiang and Sun, Chuyue Livia and Huang, Jeff and Yu, Cody Hao and Cao, Shiyi and Kozyrakis, Christos and Stoica, Ion and Gonzalez, Joseph E and others},
  journal={Advances in neural information processing systems},
  volume={37},
  pages={62557--62583},
  year={2024}
}

@misc{vllm_v0_8_2,
  title        = {vLLM(v0.8.2)},
  author       = {{vLLM Project}},
  year         = {2024},
  howpublished = {\url{https://github.com/vllm-project/vllm/tree/v0.8.2}},
  note         = {Accessed: 2026-01-27}
}

@misc{openai_chatgpt,
  title        = {ChatGPT},
  author       = {{OpenAI}},
  year         = {2024},
  howpublished = {\url{https://openai.com/zh-Hans-CN/index/chatgpt/}},
  note         = {Accessed: 2026-01-27}
}

@misc{deepseek,
  title        = {DeepSeek},
  author       = {{DeepSeek}},
  year         = {2024},
  howpublished = {\url{https://www.deepseek.com/}},
  note         = {Accessed: 2026-01-27}
}

@misc{llama_official,
  title        = {Llama},
  author       = {{Meta AI}},
  year         = {2024},
  howpublished = {\url{https://www.llama.com/}},
  note         = {Accessed: 2026-01-27}
}

@misc{distilbert_docs,
  title        = {DistilBERT},
  author       = {{Hugging Face}},
  year         = {2024},
  howpublished = {\url{https://huggingface.co/docs/transformers/model_doc/distilbert}},
  note         = {Hugging Face Transformers Documentation; Accessed: 2026-01-27}
}

@misc{cursor_ai,
  title        = {Cursor},
  author       = {{Cursor}},
  year         = {2024},
  howpublished = {\url{https://cursor.com/}},
  note         = {Accessed: 2026-01-27}
}

@misc{faiss_meta,
  title        = {FAISS: A Library for Efficient Similarity Search and Clustering of Dense Vectors},
  author       = {{Meta AI}},
  howpublished = {\url{https://ai.meta.com/tools/faiss/}},
  year         = {2024},
  note         = {Accessed: 2026-01-28}
}

@misc{qwen3_32b,
  title        = {Qwen3-32B},
  author       = {Qwen},
  year         = {2024},
  howpublished = {\url{https://huggingface.co/Qwen/Qwen3-32B}},
  note         = {Accessed: 2026-01-29}
}

@misc{llama3_1_8b_instruct,
  title        = {Llama-3.1-8B-Instruct},
  author       = {Meta AI},
  year         = {2024},
  howpublished = {\url{https://huggingface.co/meta-llama/Llama-3.1-8B-Instruct}},
  note         = {Accessed: 2026-01-29}
}

@inproceedings{gong2025past,
  title={Past-future scheduler for llm serving under sla guarantees},
  author={Gong, Ruihao and Bai, Shihao and Wu, Siyu and Fan, Yunqian and Wang, Zaijun and Li, Xiuhong and Yang, Hailong and Liu, Xianglong},
  booktitle={Proceedings of the 30th ACM International Conference on Architectural Support for Programming Languages and Operating Systems, Volume 2},
  pages={798--813},
  year={2025}
}

@article{qiu2024efficient,
  title={Efficient interactive llm serving with proxy model-based sequence length prediction},
  author={Qiu, Haoran and Mao, Weichao and Patke, Archit and Cui, Shengkun and Jha, Saurabh and Wang, Chen and Franke, Hubertus and Kalbarczyk, Zbigniew T and Ba{\c{s}}ar, Tamer and Iyer, Ravishankar K},
  journal={arXiv preprint arXiv:2404.08509},
  year={2024}
}

@inproceedings{po,
author = {Zheng, Zangwei and Ren, Xiaozhe and Xue, Fuzhao and Luo, Yang and Jiang, Xin and You, Yang},
title = {Response length perception and sequence scheduling: an LLM-empowered LLM inference pipeline},
year = {2023},
publisher = {Curran Associates Inc.},
address = {Red Hook, NY, USA},
booktitle = {Proceedings of the 37th International Conference on Neural Information Processing Systems},
articleno = {2859},
numpages = {14},
location = {New Orleans, LA, USA},
series = {NIPS '23}
}

@article{dao2022flashattention,
  title={Flashattention: Fast and memory-efficient exact attention with io-awareness},
  author={Dao, Tri and Fu, Dan and Ermon, Stefano and Rudra, Atri and R{\'e}, Christopher},
  journal={Advances in neural information processing systems},
  volume={35},
  pages={16344--16359},
  year={2022}
}

@article{wang2023voyager,
  title   = {Voyager: An Open-Ended Embodied Agent with Large Language Models},
  author  = {Wang, Guanzhi and Zhang, Yu and Chen, Han and Xiong, Hao and Wang, Wenlong and Li, Linjie and Liu, Tong and Wang, Yu and Wei, Yujie and Guo, Duyu},
  journal = {arXiv preprint arXiv:2305.16291},
  year    = {2023}
}

@article{dao2023flashattention,
  title={Flashattention-2: Faster attention with better parallelism and work partitioning},
  author={Dao, Tri},
  journal={arXiv preprint arXiv:2307.08691},
  year={2023}
}

@article{dettmers2022gpt3,
  title={Gpt3. int8 (): 8-bit matrix multiplication for transformers at scale},
  author={Dettmers, Tim and Lewis, Mike and Belkada, Younes and Zettlemoyer, Luke},
  journal={Advances in neural information processing systems},
  volume={35},
  pages={30318--30332},
  year={2022}
}

@article{frantar2022gptq,
  title={Gptq: Accurate post-training quantization for generative pre-trained transformers},
  author={Frantar, Elias and Ashkboos, Saleh and Hoefler, Torsten and Alistarh, Dan},
  journal={arXiv preprint arXiv:2210.17323},
  year={2022}
}

@inproceedings{zhong2024distserve,
  title={$\{$DistServe$\}$: Disaggregating prefill and decoding for goodput-optimized large language model serving},
  author={Zhong, Yinmin and Liu, Shengyu and Chen, Junda and Hu, Jianbo and Zhu, Yibo and Liu, Xuanzhe and Jin, Xin and Zhang, Hao},
  booktitle={18th USENIX Symposium on Operating Systems Design and Implementation (OSDI 24)},
  pages={193--210},
  year={2024}
}

@article{qin2024mooncake,
  title={Mooncake: A kvcache-centric disaggregated architecture for llm serving},
  author={Qin, Ruoyu and Li, Zheming and He, Weiran and Cui, Jialei and Tang, Heyi and Ren, Feng and Ma, Teng and Cai, Shangming and Zhang, Yineng and Zhang, Mingxing and others},
  journal={ACM Transactions on Storage},
  year={2024},
  publisher={ACM New York, NY}
}

@article{hu2024memserve,
  title={Memserve: Context caching for disaggregated llm serving with elastic memory pool},
  author={Hu, Cunchen and Huang, Heyang and Hu, Junhao and Xu, Jiang and Chen, Xusheng and Xie, Tao and Wang, Chenxi and Wang, Sa and Bao, Yungang and Sun, Ninghui and others},
  journal={arXiv preprint arXiv:2406.17565},
  year={2024}
}

@inproceedings{agrawal2024taming,
  title={Taming $\{$Throughput-Latency$\}$ tradeoff in $\{$LLM$\}$ inference with $\{$Sarathi-Serve$\}$},
  author={Agrawal, Amey and Kedia, Nitin and Panwar, Ashish and Mohan, Jayashree and Kwatra, Nipun and Gulavani, Bhargav and Tumanov, Alexey and Ramjee, Ramachandran},
  booktitle={18th USENIX Symposium on Operating Systems Design and Implementation (OSDI 24)},
  pages={117--134},
  year={2024}
}

@article{xiao2023efficient,
  title={Efficient streaming language models with attention sinks},
  author={Xiao, Guangxuan and Tian, Yuandong and Chen, Beidi and Han, Song and Lewis, Mike},
  journal={arXiv preprint arXiv:2309.17453},
  year={2023}
}

@article{zhang2023h2o,
  title={H2o: Heavy-hitter oracle for efficient generative inference of large language models},
  author={Zhang, Zhenyu and Sheng, Ying and Zhou, Tianyi and Chen, Tianlong and Zheng, Lianmin and Cai, Ruisi and Song, Zhao and Tian, Yuandong and R{\'e}, Christopher and Barrett, Clark and others},
  journal={Advances in Neural Information Processing Systems},
  volume={36},
  pages={34661--34710},
  year={2023}
}

@inproceedings{xiang2025aegaeon,
  title        = {Aegaeon: Effective {GPU} Pooling for Concurrent {LLM} Serving on the Market},
  author       = {Xiang, Yuxing and Li, Xue and Qian, Kun and Yang, Yufan and Zhu, Diwen and Yu, Wenyuan and Zhai, Ennan and Liu, Xuanzhe and Jin, Xin and Zhou, Jingren},
  booktitle    = {Proceedings of the ACM SIGOPS 31st Symposium on Operating Systems Principles (SOSP '25)},
  year         = {2025},
  doi          = {10.1145/3731569.3764815},
  numpages     = {16}
}

@inproceedings{leviathan2023speculative,
  author    = {Yaniv Leviathan and Matan Kalman and Yossi Matias},
  title     = {Fast Inference from Transformers via Speculative Decoding},
  booktitle = {International Conference on Machine Learning (ICML)},
  series    = {Proceedings of Machine Learning Research},
  volume    = {202},
  pages     = {19274--19286},
  publisher = {PMLR},
  year      = {2023},
  url       = {https://proceedings.mlr.press/v202/leviathan23a.html}
}

@article{miao2023specinfer,
  author     = {Xupeng Miao and Gabriele Oliaro and Zhihao Zhang and Xinhao Cheng and Zeyu Wang and Rae Ying Yee Wong and Zhuoming Chen and Daiyaan Arfeen and Reyna Abhyankar and Zhihao Jia},
  title      = {SpecInfer: Accelerating Generative {LLM} Serving with Speculative Inference and Token Tree Verification},
  journal    = {CoRR},
  volume     = {abs/2305.09781},
  year       = {2023},
  doi        = {10.48550/ARXIV.2305.09781},
  url        = {https://doi.org/10.48550/arXiv.2305.09781},
  eprinttype = {arXiv},
  eprint     = {2305.09781}
}

@article{liu2023online,
  author     = {Xiaoxuan Liu and Lanxiang Hu and Peter Bailis and Ion Stoica and Zhijie Deng and Alvin Cheung and Hao Zhang},
  title      = {Online Speculative Decoding},
  journal    = {CoRR},
  volume     = {abs/2310.07177},
  year       = {2023},
  doi        = {10.48550/ARXIV.2310.07177},
  url        = {https://doi.org/10.48550/arXiv.2310.07177},
  eprinttype = {arXiv},
  eprint     = {2310.07177}
}

@inproceedings{liu2024kivi,
  author    = {Zirui Liu and Jiayi Yuan and Hongye Jin and Shaochen (Henry) Zhong and Zhaozhuo Xu and Vladimir Braverman and Beidi Chen and Xia Hu},
  title     = {{KIVI:} {A} Tuning-Free Asymmetric 2bit Quantization for {KV} Cache},
  booktitle = {International Conference on Machine Learning (ICML)},
  publisher = {OpenReview.net},
  year      = {2024},
  url       = {https://openreview.net/forum?id=L057s2Rq8O}
}

@inproceedings{duan2024muxserve,
  author    = {Jiangfei Duan and Runyu Lu and Haojie Duanmu and Xiuhong Li and Xingcheng Zhang and Dahua Lin and Ion Stoica and Hao Zhang},
  title     = {MuxServe: Flexible Spatial-Temporal Multiplexing for Multiple {LLM} Serving},
  booktitle = {International Conference on Machine Learning (ICML)},
  publisher = {OpenReview.net},
  year      = {2024},
  url       = {https://openreview.net/forum?id=R0SoZvqXyQ}
}

@inproceedings{sun2024llumnix,
  author    = {Biao Sun and Ziming Huang and Hanyu Zhao and Wencong Xiao and Xinyi Zhang and Yong Li and Wei Lin},
  title     = {Llumnix: Dynamic Scheduling for Large Language Model Serving},
  booktitle = {18th {USENIX} Symposium on Operating Systems Design and Implementation (OSDI)},
  pages     = {173--191},
  publisher = {USENIX Association},
  year      = {2024},
  url       = {https://www.usenix.org/conference/osdi24/presentation/sun-biao}
}

@inproceedings{lee2024infinigen,
  author    = {Wonbeom Lee and Jungi Lee and Junghwan Seo and Jaewoong Sim},
  title     = {InfiniGen: Efficient Generative Inference of Large Language Models with Dynamic {KV} Cache Management},
  booktitle = {18th {USENIX} Symposium on Operating Systems Design and Implementation (OSDI)},
  pages     = {155--172},
  publisher = {USENIX Association},
  year      = {2024},
  url       = {https://www.usenix.org/conference/osdi24/presentation/lee}
}

@inproceedings{ousterhout2015making,
  title={Making sense of performance in data analytics frameworks},
  author={Ousterhout, Kay and Rasti, Ryan and Ratnasamy, Sylvia and Shenker, Scott and Chun, Byung-Gon},
  booktitle={12th USENIX Symposium on Networked Systems Design and Implementation (NSDI 15)},
  pages={293--307},
  year={2015}
}

@inproceedings{lozi2016linux,
  title={The Linux scheduler: a decade of wasted cores},
  author={Lozi, Jean-Pierre and Lepers, Baptiste and Funston, Justin and Gaud, Fabien and Qu{\'e}ma, Vivien and Fedorova, Alexandra},
  booktitle={Proceedings of the Eleventh European Conference on Computer Systems},
  pages={1--16},
  year={2016}
}

@inproceedings{abrossimov1989generic,
  title={Generic virtual memory management for operating system kernels},
  author={Abrossimov, E and Rozier, Marc and Shapiro, Marc},
  booktitle={Proceedings of the twelfth ACM symposium on Operating systems principles},
  pages={123--136},
  year={1989}
}

@incollection{mahajan2008multi,
  title={Multi-armed bandit problems},
  author={Mahajan, Aditya and Teneketzis, Demosthenis},
  booktitle={Foundations and applications of sensor management},
  pages={121--151},
  year={2008},
  publisher={Springer}
}

@inproceedings{cai2025r,
  title={R-kv: Redundancy-aware kv cache compression for reasoning models},
  author={Cai, Zefan and Xiao, Wen and Sun, Hanshi and Luo, Cheng and Zhang, Yikai and Wan, Ke and Li, Yucheng and Zhou, Yeyang and Chang, Li-Wen and Gu, Jiuxiang and others},
  booktitle={The Thirty-ninth Annual Conference on Neural Information Processing Systems},
  year={2025}
}
